%% file: main.tex
\documentclass[journal,transmag]{IEEEtran}
\usepackage{graphicx}
\usepackage{soul}

\usepackage{amsmath,bbm}
\usepackage{cleveref}
\usepackage{verbatim}
\usepackage{comment}
\usepackage{cases}
\usepackage{soul}
\usepackage{amssymb}
\usepackage[inline]{enumitem}
\crefrangelabelformat{}{#3#1#4 to~#5#2#6}
\crefrangeformat{equation}{(#3#1#4) to~(#5#2#6)}
\usepackage{algorithm}% http://ctan.org/pkg/algorithms
\usepackage{algpseudocode}% http://ctan.org/pkg/algorithmicx
\usepackage{color}
\usepackage{multirow}
\usepackage{tagging}
\usepackage{subcaption,xcolor}
\usepackage{listings}

\usepackage{soul}
\usepackage{tabularx}
\usepackage{booktabs}
\newcommand{\ra}[1]{\renewcommand{\arraystretch}{#1}}
\usepackage{makecell}

\newcommand{\squeezeup}{\vspace{-3mm}}

\definecolor{eclipseBlue}{RGB}{42,0.0,255}
\definecolor{eclipseGreen}{RGB}{63,127,95}
\definecolor{eclipsePurple}{RGB}{127,0,85}
\definecolor{pythonOrange}{RGB}{227,98,9}
\definecolor{pythonPurple}{RGB}{111,66,193}
\lstdefinelanguage{P4}
{
	alsoletter=\#,
	morekeywords={
		header\_type,
		fields,
		header,
		parser,
		extract,
		return,
		default,
		next,
		\#define,
		latest,
		select,
		apply,
		control,
		table,
		reads,
		actions,
		modify\_field,
		action,
		valid,
		nop,
		exact,
		lpm,
		if,
		hit,
		and,
		extern,
		void,
		bit,
		transition,
		state,
		accept
	},
	sensitive=true, % keywords are not case-sensitive
	morecomment=[l]{//}, % l is for line comment
	morecomment=[s]{/*}{*/}, % s is for start and end delimiter
	morestring=[b]" % defines that strings are enclosed in double quotes
}

\usepackage[colorinlistoftodos,textsize=tiny,textwidth=15mm]{todonotes}
\lstdefinelanguage{Python2}{
	sensitive = true,
	keywords={ReductionFunction, TriggerFunction, ActivityFunction, TimeError, AbsoluteError, None, Result, State, Device, append, from, import, as, for, in, getIngressPort, getPorts, getEdgeRouters, def, return, getLeafSwitches, getSpineSwitches, len, range, sum, and, ValueError, TimeObsolescence,UpdateError},
	otherkeywords={% Operators
		>, <, ==
	},
	keywords = [2]{states, operation, s0, obsolescenceLevel, trigger, target, activity, action, scope, filter, port, match, args, inconsistencyLevel},
	keywordstyle=\color{pythonPurple},
	keywordstyle=[2]\color{pythonOrange},% for example
	morecomment=[l]{\#},
	morestring=[b]", % defines that strings are enclosed in double quotes
}
\begin{document}

%\begin{frontmatter}

\title{LOcAl DEcisions on Replicated States (LOADER) in programmable dataplanes: programming abstraction and experimental evaluation}

%\author[1]{German Sviridov\corref{cor1}}
%\author[2]{Marco Bonola\corref{cor1}}
%\author[2]{Angelo Tulumello\corref{cor1}}
%\author[1]{Paolo Giaccone\corref{cor1}}
%\author[1]{Andrea Bianco\corref{cor1}}
%\author[2]{Giuseppe Bianchi\corref{cor1}}

\author{\IEEEauthorblockN{
                German Sviridov\IEEEauthorrefmark{1}, Marco Bonola\IEEEauthorrefmark{2}, Angelo Tulumello\IEEEauthorrefmark{2}, Paolo Giaccone\IEEEauthorrefmark{1}, Andrea Bianco\IEEEauthorrefmark{1}, Giuseppe Bianchi\IEEEauthorrefmark{2}}
\\\IEEEauthorblockA{\IEEEauthorrefmark{1} Politecnico di Torino, Torino, Italy,\hspace{1cm}
                \IEEEauthorrefmark{2}University of Rome Tor Vergata, Rome, Italy}
}

%\address[1]{Politecnico di Torino, Torino, Italy}
%\address[2]{Universit\`a degli Studi di Roma Tor Vergata, Rome, Italy}
%\author{\IEEEauthorblockN{
%		German Sviridov\IEEEauthorrefmark{1}, Marco Bonola\IEEEauthorrefmark{2}, Angelo Tulumello\IEEEauthorrefmark{2}, Paolo Giaccone\IEEEauthorrefmark{1}, Andrea Bianco\IEEEauthorrefmark{1}, Giuseppe Bianchi\IEEEauthorrefmark{2}}
%	\\\IEEEauthorblockA{\IEEEauthorrefmark{1} Politecnico di Torino, Torino, Italy,\hspace{1cm}
%		\IEEEauthorrefmark{2}University of Rome Tor Vergata, Rome, Italy}}

\maketitle

%\IEEEpeerreviewmaketitle

\input{abstract}
%end{frontmatter}

\input{intro}

\input{related}

\input{overview}

\input{abstraction}
\input{lodge_implementation}

\input{mapping}

\input{other_applications}

\input{conclusion}

\bibliography{biblio}
\input{appendix_codes}

\end{document}

%% file: abstract.tex
%!TEX root = main.tex
%!TEX spellcheck = en_US

\begin{abstract}

Programmable data planes recently emerged as a prominent innovation in Software Defined Networking (SDN). They provide support for stateful per-packet/per-flow operations over hardware network switches specifically designed for network processing. Unlike early SDN solutions such as OpenFlow, modern stateful data planes permit to keep (and dynamically update) per-flow states local to each switch, thus dramatically improving reactiveness of network applications to different state changes. Still, in stateful data planes, the management of non-local states is assumed to be completely delegated to a centralized controller, thus requiring extra overhead to be accessed. 

Our LOADER proposal aims at contrasting the apparent dichotomy between local and non-local states. We do so by introducing a {\em new} possibility: permit to take localized (in-switch) decisions not only on local states but also on global replicated  states, thus providing support for network-wide applications without incurring the drawbacks of classic approaches. To this purpose, i) we provide high-level programming abstractions devised to define the states and the update logic of a generic network-wide application, and ii) we detail the underlying low level state management and replication mechanisms. We then show LOADER's independence of the stateful data plane technology employed, by implementing it over two distinct stateful data planes (P4 switches and OPP - Open Packet Processor - switches), and by experimentally validating both implementations in an emulated testbed using a simple distributed Deny-of-Service (DoS) detection application. 
\end{abstract}

%% file: intro.tex
%!TEX root = main.tex
%!TEX spellcheck = en_US

\section{Introduction}
Future networks are called to efficiently and flexibly support an ever growing variety of heterogeneous network functions such as network address translation, tunneling, load balancing, traffic engineering, monitoring, intrusion detection, and so on. Software-based programmability of such type of functions has been first pioneered by early Software Defined Networking (SDN) proposals, and then by the more recent trend of Network Function Virtualization (NFV). However, both these approaches have shown shortcomings. Indeed, original SDN approaches (and, more specifically, the OpenFlow-based ones), were relying on stateless switching architectures, and thus suffered of the need to centralize {\em any} state update and maintenance to a centralized controller, thus paying a significant toll in terms of latency and communication overhead. On the other side, NFV has addressed the design of middlebox functionalities in software, typically using commodity CPUs. However, early NFV implementations appeared to be performance-limited: it is a fact that there exists a substantial gap  (a $50\times$ factor) between the speed attainable in software opposed to dedicated HW devices, and such gap is not going to decrease in the future, with HW switches capable to attain many Terabit per seconds, opposed to the tens of Gigabit per second attainable by their SW counterparts.

In order to overcome such limitations, starting from 2014 with OpenState~\cite{openstate14} and P4~\cite{bosshart2014p4}, a new innovation trend emerged with the introduction of {\em programmable} / {\em stateful} data planes. Stateful data planes offer an additional level of programmability with respect to the traditional stateless SDN paradigm, by introducing the possibility of keeping and manipulating persistent states locally at the network device. Opposed to stateless switches, persistent states can now be directly deployed and managed inside network devices in the form of simple user-defined memory elements. Furthermore, arbitrary algorithms for packet/flow processing, e.g., described in terms of simple Mealy Finite State Machines~\cite{openstate14} or more sophisticated Extended Finite State Machines~\cite{OPP,flowbraze}, can be directly loaded and run inside the processing pipeline of individual network devices, thus providing opportunities of implementing network applications directly within the network device at line rate.

The crucial advantage of stateful data plane technologies consists in the possibility to significantly reduce the interaction between switches and the controller. Opposed to a stateless data plane, in which any change of the forwarding decision requires the intervention of the controller, a stateful data plane permits to take {\em localized} decisions, i.e., adapt the forwarding behavior to network events and handle changing states locally 
%\st{state changes locally happening (and handled)} 
inside the switch.  This approach significantly reduces the reliance upon a centralized controller, and mitigates the relevant severe penalties in terms of latency and signaling overhead~\cite{yeganeh2013scalability}, hence greatly improving the reactivity of network control applications. 

%\todo{TODO: AGGIUNGERE CONCETTO DELL'ABSTRACT: O LOCAL O GLOBAL: E' POSSIBILE TROIVARE UNA TERZA VIA?}

%Unfortunately, the beneficial effects of distributing network applications on stateful switches cannot be achieved for non-local states. 
Unfortunately, the benefits of distributing network applications on stateful switches cannot be achieved in cases where non-local states need to be considered. For example, an application that identifies the occurrence of a particular event based on  multiple statistics gathered from different switches, operates on a global state that is the combination of different local statistics of different switches. Even in the case of stateful data planes, the control and update of the global state is still delegated to a centralized entity, either to a controller or a single switch~\cite{snap16}. The traditional approach of employing a centralized controller for global state management greatly simplifies the implementation, but non-local states can be accessed and updated only at the price of extra delay, thus affecting the overall reactivity. % of network control applications.
%Recent proposals for stateful SDN have been assuming a single replica of each state, stored in one single switch across the whole network. This leads to possible scalability issues in the case of {\em states} shared across multiple network applications or depending on global network dynamics. Indeed, 
On the other hand, solutions employing global states centralized in a stateful switch lead to performance impairments. Indeed, all flows affected by/ affecting a global state should traverse the switches storing it. This ultimately leads to an overall higher network utilization and traffic concentration, thus affecting network congestion and  available capacity. Furthermore, any failure to the switch can jeopardize the state integrity due to the presence of a single replica of the global state.
%: because  any given state is unique inside the network, the failure of a network device will lead to an irreversible loss of the value of the states stored within the device. 

In this work we propose a novel framework, namely LOADER (LOcAl DEcisions on Replicated states), which enables a new possibility for stateful data planes: the states and the corresponding control logic are distributed across the switches and the controller, while permitting multiple replicas of the same state/control logic to be present  in the network. This permits to run network applications operating on global states without a unique central entity. Switches can take instantaneous decisions based on local replicas of non-local states, without any controller intervention, thus re-establishing the beneficial effects of stateful data planes also for non-local states.
%
%The main goals of LOADER are: (i) to support replicated states;
%reduce the burden for the SDN controller in managing network-wide applications; 
%(ii) to provide better utilization of the underlying network computational resources; (iii) to improve application  reactiveness by removing the latency introduced by the communication with the controller. 
%These goals are achieved thanks to:
LOADER provides:
\begin{itemize} 
\item the programming abstractions to define generic (either local or non-local) states and the control logic of any network application;
\item the engine to optimally embed the states and the control logic into the network devices and the controller, to optimize performance while taking into account the available resources in terms of processing and state storage capabilities;
\item the mechanism to transparently replicate non-local states across multiple network devices. 
% This mechanism permits network devices to continuously share their local view of the network among each other and, as a consequence, to take local decisions based on either the readily available local state or based on network-wide states received by other network devices.
\end{itemize}
%In our work, we present a lightweight implementation of LOADER inside ONOS~\cite{berde2014onos} controller and we provide an experimental validation of the whole framework based on P4~\cite{bosshart2014p4} and OPP~\cite{OPP} stateful data planes by running some example applications.

The rest of the paper is organized as follows. In Sec.~\ref{sec:related} we discuss the related work.
In Sec.~\ref{sec:offloading_applications} we discuss the issues and possible solutions for offloading network applications to the data plane. 
In Sec.~\ref{sec:LOADER_programming_model} we first provide a high level abstraction of the LOADER framework by defining its core modules and later delve into the details of each module and the way LOADER abstraction is exposed to the network programmer. 
In Sec.~\ref{sec:consistency_analysis} we analyze consistency-related issues when dealing with replicated states and how to overcome them. In Sec.~\ref{sec:LOADER_implementation} we describe how we implemented a lightweight version of the LOADER framework in ONOS~\cite{berde2014onos} with major emphasis on the data plane implementation in P4~\cite{bosshart2014p4} and Open Packet Processor (OPP)~\cite{OPP}. The implementation on these two different architectures is aimed at showing the generality of the proposed programming model, which is agnostic of the adopted data plane implementation.
In Sec.~\ref{sec:ddos} we show how to program a distributed Deny-of-Service (DoS) detection application in LOADER and experimentally assess the performance for both P4 and OPP based implementations. Furthermore, to highlight the versatility of the proposed framework, we provide details about the implementation of other network applications. The corresponding LOADER code is reported in the appendix. 
Finally, we draw our conclusions in Sec.~\ref{sec:conc}.

%% file: related.tex
%!TEX root = main.tex
%!TEX spellcheck = en_US
\section{Related work}\label{sec:related}
Data plane embedding of network applications is steadily gaining attention from the industry and the research community. Numerous frameworks and abstraction models have been proposed which try to expose to the programmer data plane resources allowing them to embed custom logic and persistent states directly inside the data plane. Yet no significant effort have been put into dealing with scalability issues which inevitably arise when embedded network applications must operate on a network-wide states.

Numerous studies considered employing replicated states in the data plane~\cite{alizadeh2014conga, raghavan2007cloud, harrison2018network}, demonstrating the scalability benefits of such approach and treating it as an enabler for new network applications. Yet all of the above studies considered specific applications with tailored implementation.

On the other hand there have been studies~\cite{kim2015kinetic, yuan2014netegg, beckett2016temporal, yeganeh2016beehive, mcclurg2016event} concerning the development of general purpose programming abstractions for network applications, none of which considered having replicated states in the data plane.
In particular, in~\cite{kim2015kinetic, yuan2014netegg, beckett2016temporal} the authors tried to address the issues of defining a general enough programming language for network applications. Yet they considered that states are kept at the controller in a centralized fashion, thus not only neglecting the available data plane resources but also leading to scalability issues due to the centralization of all policies at the same controller.
In~\cite{yeganeh2016beehive} the authors addressed the former issue by providing an abstraction model including replicated states and distributed network applications among different controllers. Although solving the issue with scalability, applications still reside in the control plane, mitigating the benefits of having stateful data planes.

On the contrary, works such as~\cite{snap16,mcclurg2016event} proposed novel network programming abstractions, which permits to define complex network applications for stateful data planes. In particular, SNAP~\cite{snap16} addressed the problem of performing optimal embedding of states across the network switches, taking into account the dependency between states and the traffic flows. 
Nevertheless, by design, SNAP is limited to just one replica of each state within the network, thus still precluding a wide variety of possible network applications.

%In~\cite{zeineddine2018stateful} the authors focus on providing state redundancy and traffic load balancing by employing independent copies of the same application. LOADER is able to achieve the same goal by employing state replication instead of performing full application copy, leading to better hardware resource utilization.
LOADER, instead, enables multiple replicas of the state, extending the single replica approach of SNAP.
Furthermore, LOADER closes the gaps of previously proposed programming models by providing a programmer-level abstraction for the definition of network applications while transparently dealing with replication and embedding problem.

%Swing State~\cite{luo2017swing} introduces a mechanism which provide state migrations entirely in the data plane but, as in the case of SNAP, assumes only one replica of each state, which is migrated across the network. 

The optimal replication problem for multiple replicas has been defined and investigated in~\cite{abu2019}. Given a network application and the corresponding states, the problem considers all the traffic flows that are affected by/affect such states and, based on a  generic cost function, computes (i) the  optimal number of replicas, (ii) their placement within the network and (iii) the corresponding optimal traffic routing.  The work in~\cite{abu2019} can be used as a building block for LOADER (i.e., the optimization engine), which provides the programming framework and the  implementation for replicated states.

%Stateful NetKAT \cite{mcclurg2016event} is a programming abstraction for the development of network applications. Differently from SNAP, NetKAT provides a native support for replicated states, yet by design the actual state replication can be performed only at the edges of the network. Moreover, differently from LOADER, the traffic affected by/affecting the replicated states is constrained to traverse all replicas, thus precluding a wide range of applications.

In general, the problem of maintaining consistency across replicated states has been deeply investigated in the past in the field of distributed systems~\cite{ozsu2011principles} and many solutions have been proposed, depending on the nature of the states, the desired properties and the available resources. %In LOADER, we leverage the proposed approaches to achieve consistency between replicas of the same state by considering the limitations in terms of resources of stateful switches.
There have been however, few works concerning replication in stateful data planes. %Among them NetPaxos~\cite{netpaxos} provides application-layer acceleration for Paxos~\cite{paxos} consensus protocol by offloading parts of the algorithm to the switches. 
Although we do not treat the issue of developing a sophisticated replication algorithm in this paper, we design LOADER in an agnostic way to the actual consistency scheme in view of the future research in the field.
%Differently from NetPaxos, LOADER provides state replication directly in the data plane.

A preliminary version of this work was presented in~\cite{netsoft}, focusing on some implementation issues of LOADER and providing some experimental results. Furthermore, \cite{netsoft} did not consider the abstraction model required to develop network applications based on replicated states.

%The problem of obtaining a global view locally at each device is well known in the literature and it is usually referred to as {\em Global Predicate Evaluation} (GPE) \cite{babaoglu1993consistent}. LOADER takes inspiration from GPE in the definition of its abstraction model. The proposed abstraction provides the definition of the basic elements of the global predicates, which in this case is referred to as network-wide application, and their mapping inside the network from both physical and logical point of view. 
%Our goal is to provide zero processing delay for state replication replication in order to maintain high system reactiveness in the presence of critical events.
%\begin{itemize}
%%\item P4
%\item NetPaxos \cite{netpaxos}
%%\item E-State \cite{peuster2016state} Distributed states in NFV but with on demand pulls. 
%%\item Swing State \cite{luo2017swing}
%%\item VNE \cite{fischer2013virtual}
%\end{itemize}

%% file: overview.tex
%!TEX root = main.tex
\section{Offloading network applications}\label{sec:offloading_applications}
Classic SDN management schemes present a series of limitations such as poor reactiveness, big communication overhead and compromised fault-tolerance caused by the excessive centralization of the control plane. Stateful data planes introduce the possibility of embedding custom logic inside network devices, thus offering a new way of mitigating the aforementioned issues by providing means of {\em offloading control plane functionalities to the data plane}.

\subsection{Network application in stateless SDN}
In traditional SDN networks a logically centralized entity, namely the controller, is responsible for managing the whole network operations by means of user-defined network applications.

Being centralized, to function correctly, network applications are required to operate on an accurate snapshot of the network. The task of constructing such snapshot is delegated to the controller which continuously gathers network statistics in the form of network {\em states}, which provide a synthetic description of the network in the form of generic data structure holding a variable or a compound of variables.
Given this information at the controller, applications are able to detect the presence of certain events (e.g., load unbalance, security risks, misconfiguration, etc.) by performing a set of operations over the states and, whenever possible, take actions to correct the network operations.

\subsection{Network applications offloading with stateful data planes}
Although being suitable for coarse-grained network operations, due to the poor reactiveness, classic SDN approaches come to their limits when it comes to supporting network applications performing fine-grained operations. Such is the case of, e.g., per-packet processing or fine-grained traffic engineering~\cite{he2015measuring}.
Stateful data planes offer the possibility of mitigating this limitation by offloading the related logic to the data plane.

Offloading network applications implies the embedding of some or all of the application elements into the network devices. This involves application states being embedded under the form of stateful primitives natively supported by the network devices and action logic under the form of data plane packet processing modules. Although being feasible from the theoretical point of view, application offloading creates considerable challenges in practice.

\subsection{Satisfying resource constraints}
When it comes to offloading, the type and the corresponding amount of available resources at each network device pose hard constraints for the embedding of application elements.

Dedicated hardware devices, such as switches and routers, lead to almost zero latency during the execution of local processing, but typically have limited resources in terms of processing capabilities and memory. On the other hand, general purpose network devices such as SDN controllers provide resource flexibility at the cost of large processing latency.
To minimize the application execution latency, during the embedding phase, network applications exceeding the resources constraint at a single network device may be be \emph{split across multiple devices}. If application splitting still does not satisfy the resources constraint, the application may be fully delegated to the controller, thus, reverting to a traditional stateless SDN scheme.

\subsection{Inter and intra application dependencies}
In addition to the resource constraints, most of the applications involve a dependency among different elements of a network application, i.e., states are accessed/modified and actions are executed according to a well-defined order which is tightly bound to the definition of the application. 
The complexity is further increased when considering that a given state of an application may be accessed by different network applications such as in the case of two network applications reading a common counter.
This inter and intra application dependency imposes a constraint on how the traffic must be routed across individual elements of the split application to ensure the correctness on the execution of the application~\cite{snap16,abu2019}. 

\begin{figure}[!tb]
	\centering
	\begin{subfigure}{.5\columnwidth}
		\centering
		\includegraphics[height=75pt]{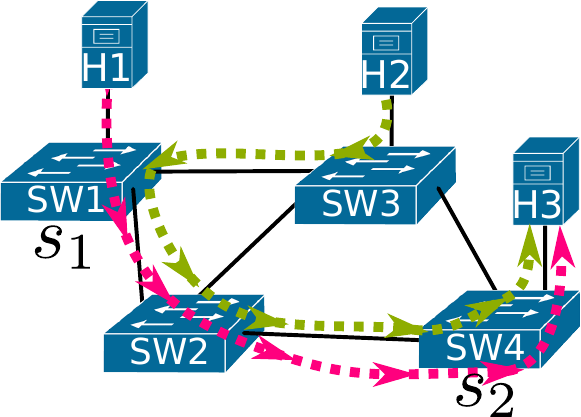}
		\label{fig:snap_routing}
	\end{subfigure}%
	\begin{subfigure}{.5\columnwidth}
		\centering
		\includegraphics[height=75pt]{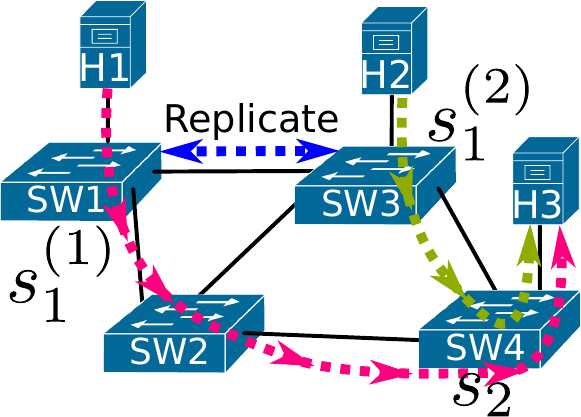}
	\end{subfigure}
	\caption{Example routing without replicated states (left) and with replicated states (right), as enabled by LOADER. }
	\label{fig:reduction_decopmosition}
\end{figure}

\subsection{Offloading shared states}
Considering a general case of network applications embedded in different network devices we can define two macro-categories of states: 
i) given a generic state $s$ stored in a given network device $n$, $s$ is said to be {\em local} if it can be accessed (read/write) only by $n$ itself. In such scenario, $s$ can be internally embedded in $n$ (provided that $n$ is capable of supporting it). 
ii) On the contrary, when $s$ is accessed (read/write) by multiple network devices that share the state, $s$ is said to be non-local.
If all states related to a network application are local, the offloading does not present any considerable challenge as states can be embedded into a single network device, assuming no violation of the capacity constraint. However, when a state is non-local, multiple network applications or multiple parts of the same application must be able to access the state. 
%Following this, a non-local state can be embedded in different network devices related to either different applications or different components of the same application and a synchronization scheme must operate among all the replicas.

In classic stateless SDN, non-local states are managed by states polling and aggregation at the controller. Instead, in stateful SDN,  non-local states can be supported with one of the two approaches:
\begin{itemize}[leftmargin=*]
\item {\em Single replica}. As proposed in SNAP~\cite{snap16}, a non-local state can be embedded in a single network device, thus a unique replica is made available in the entire network. Consequently, to support inter/intra application dependency all traffic affected by/affecting the state must be routed through that device.
In SNAP, the choice of the network device to embed the state in is optimized according to some optimization criteria, e.g., minimization of the distance among dependent states, equal load-balancing across the network devices, etc. 

The approach proposed by SNAP may lead to major scalability and performance impairments, specifically when a state is affected by/affects a large amount of traffic.
\item {\em Multiple replicas}: 
A single state is made available in different parts of the network by providing copies (i.e., replicas) of it inside different network devices. This approach permits to distribute the traffic across multiple network devices while also providing robustness to failures. However, although this approach provides more embedding flexibility, it requires the presence of a replication protocol between the replicas, to keep all replicas consistent. In the absence of such replication mechanism, the values of each replicated state will start to diverge, thus leading to different distinct states which will not be representative anymore of the global network dynamics.
\end{itemize}

An example of the the two approaches is depicted in Fig.~\ref{fig:reduction_decopmosition}. Assume a network application composed of two states, namely $s_1$ and $s_2$, and two flows originating from \verb|H1| and \verb|H2| and directed towards \verb|H3|. For a single replica in \verb|SW1|, the green flow is forced to make a detour from its shortest path to traverse \verb|SW1| storing $s_1$.
On the contrary, in the presence of multiple replicas the green flow can reach its destination following the shortest path thanks to the presence of two replicas of $s_1$, namely $s_1^{(1)}$ and $s_1^{(2)}$, embedded respectively inside \verb|SW1| and \verb|SW2|.
Although being a simple example, it highlights the importance of using replicated states. Detouring flows from their shortest path adds a considerable data overhead in the network which in turns leads to ineffective use of network resources. Furthermore, in extreme cases, such as sudden traffic spikes, this resource mismanagement may lead to scenarios of excessive overload of network devices storing the state, thus degradating the flow performance. 
State replication mitigates these issues by providing multiple copies of the same state which, in the best case scenario, are all located on the shortest path for each flow. Furthermore, flow processing and updating network states are delegated to multiple switches, thus the overall workload is distributed across the switches.
\subsection{Managing inconsistency of replicated states}
The management of state inconsistencies is among the most challenging aspect of the approach employing state replication. Whenever a given replica of a state propagates its update to other replicas a period of inconsistency is created. During this time interval read operations on different replicas of the same state may lead to different outcomes.
When developing the application, the programmer must be able to take into account the presence of these errors and specify the maximum amount of error that can be tolerated. Consequently, operating on replicated states requires an additional abstraction layer capable of translating user-defined consistency constraints into embedding constrains.

In addition of defining a formal model for managing inconsistency errors, LOADER provides a general abstraction model and a framework for developing network applications based on replicated states. 
In the following, we identify a common abstraction for network applications permitting LOADER to be target independent and completely agnostic to the underlying network hardware. 
The abstraction is made generic by: 
i) supporting network applications operating only on local states, as they fall into the special-case category of single-replica states,
ii) supporting the absence of stateful switches,
iii) being target-independent from the technologies employed in the data plane.

%% file: abstraction.tex
%!TEX root = main.tex
%!TEX spellcheck = en_US

%\section{Offloading Network-Wide applications}
\section{LOADER abstraction model and framework}\label{sec:LOADER_programming_model}

\begin{figure}[!tb]
	\centering
	\includegraphics[width=.85\columnwidth]{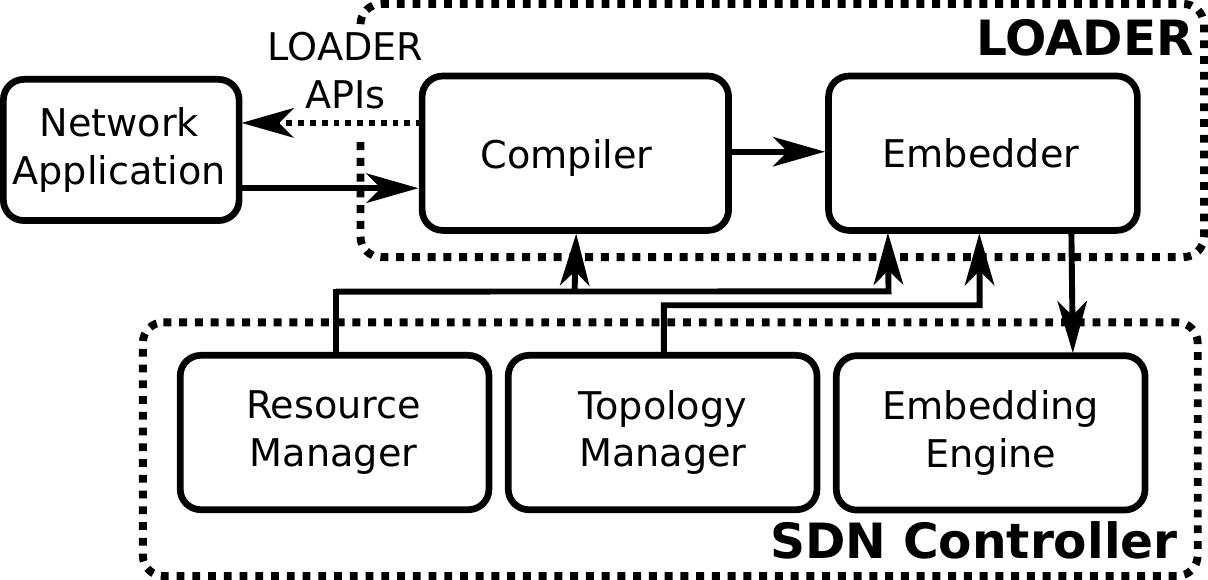}
	\caption{Main building blocks of LOADER framework.}
	\label{fig:LOADER_framework}
\end{figure}

LOADER  naturally extends functionalities of previously proposed frameworks based on single-replica states.
As shown in Fig.~\ref{fig:LOADER_framework}, the proposed framework is based on three main blocks which define the lifecycle of deploying a LOADER application: 
i) application are defined by means of a predefined set of APIs which expose to the programmer LOADER-specific functionalities;
ii) once defined, the applications undergo a compilation phase by means of a compiler capable of translating them into basic primitives supported by network devices;
iii) finally, the compiled network applications undergo an embedding phase during which the embedder will try to place the basic primitives composing the network application inside the available network devices.

In the following section we define an abstraction model for LOADER that permits the decomposition of a network application in basic elements that can be directly embedded into network devices.

\subsection{Application definition}\label{sec:application-definition}

At the top layer, users define network applications by employing a set of predefined building blocks, namely {\em application elements}, in a completely agnostic way with respect to the remaining components of the framework. The application elements supported by LOADER are the only part of the framework exposed to the programmer by means of APIs and generic language libraries.
While maintaining generality, the use of these elements permit an efficient decomposition of user-defined applications during the compilation phase and provide a comprehensive abstraction for the compiler during their translation to device-specific primitives. 

\begin{figure}[!tb]
	\centering
	\includegraphics[width=.85\columnwidth]{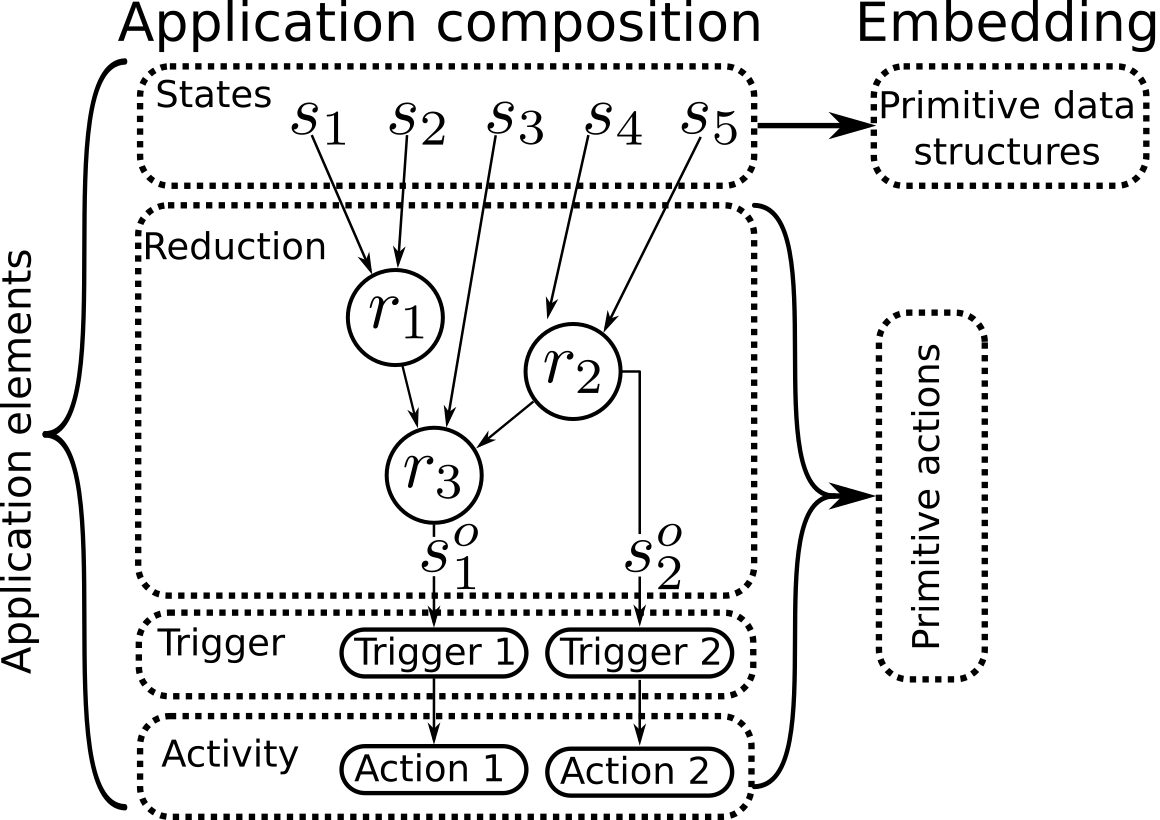}
	\caption{DAG representation of a LOADER network application and its mapping to primitive elements.}
	\label{fig:LOADER_decomposition}
\end{figure}

\subsection{Building blocks of a network application}
Fig.~\ref{fig:LOADER_decomposition} depicts an example of a generic network application employing LOADER abstraction. Each application is composed of four main types of application elements which have to be implemented by the programmer: \emph{states}, \emph{reduction functions}, \emph{trigger functions} and \emph{activity functions}.

For ease of explanation, in the following we will present the role of each application element by considering a reference data center load-balancing application. This application works as follows: (1) whenever the load on the data center servers is medium-low, the application distributes the user's request among the available servers in a load-balancing fashion, i.e., an arriving request is forwarded to the least loaded server, in terms of CPU utilization; (2) otherwise, when the data center is highly loaded, users' requests are sent to the controller for further processing.

\begin{figure}[!tb]
	%\begin{center}
	%\begin{minipage}{0.9\columnwidth}
	\begin{lstlisting}[linewidth=0.9\columnwidth, language=Python2, caption=Resource-aware load balancing with LOADER., label=code:resource-lb-initial]
	from Controller import TopologyManager
	from LOADER.PrimitiveActions import SetEgress, Rate
	from LOADER.Scope import Pkt, ExtScopeHelper
	
	THR = 0.8  # threshold CPU load percentage
	
	# Get the average CPU load of servers in the form of a list of states. We omit the details.
	loads = ExtScopeHelper(scope="ServerLoad")
	
	r1 = ReductionFunction(
				states = [loads]
				operation=argmin([i.Value() for i in loads]))
	
	r2 = ReductionFunction(
				states = [loads]
				operation=mean([i.Value() for i in loads]))
	
	a1 = ActivityFunction(
				scope = Pkt(filter = (TCP.Flag.SYN == 1)), 
				action = SetEgress,
				args = r1.Result())
	
	a2 = ActivityFunction(
				scope = Pkt(filter = (TCP.Flag.SYN == 1)),
				action = SetEgress,
				args = CONTROLLER_PORT)
	
	tr1 = TriggerFunction(
				s0=r2.Result(),
				trigger=(r2.Result() <= THR),
				inconsistencyLevel=UpdateError(15),
				activity=a1)
	
	tr2 = TriggerFunction(
				s0=r2.Result(),
				trigger=(r2.Result() > THR),
				inconsistencyLevel=UpdateError(15),
				activity=a2)
				
	\end{lstlisting}\squeezeup
	%\end{minipage}
	%\end{center}
	\end{figure}

Each application element is defined as follows:

\begin{itemize}

\item  \textbf{States:}
Let $\Omega_P=\{s_i\}_i$ be the set of states associated with a network application $P$, with $s_i^{(k)}$ be the $k$-th replica of state $s_i$, with $k\in\mathbb{N}$. For the reference load-balancing application, state $s_i$ represents the current CPU load of a generic server $i$, where $i=1,\ldots, n$,  and $n$ is the number of available servers. 

\item \textbf{Reduction function:}
The reduction function is a generic multivariate function that maps states in $\Omega_P$ to a reduced version $s^o_1$ of the input states. 
It is obtained by combining a set $\mathcal{R}=\{r_j\}$ of primitive reduction actions natively available  in the network device. In the reference application, $\mathcal{R}=\{r_1,r_2\}$ with $r_1=\arg\min()$ and $r_2=\text{mean}()$, which compute the index corresponding to the minimum and the average of an array of values,  respectively. Consequently, the reduced versions are just two scalars:  $s^o_1=\arg\min(s_1,\ldots,s_n)$ and $s^o_2=\text{mean}(s_1,\ldots,s_n)$.

\item \textbf{Trigger function:}
Based on $s^o_i$, the trigger function evaluates the presence of a particular event and decides whether a reaction is required or not.
The reference application operates concurrently on two trigger functions leading to different activity functions. The first trigger function checks if the average data center load $s^o_2$ is below a given threshold (corresponding to a low load scenario). The second trigger function instead is activated whenever $s^o_2$ is above the predefined threshold.

\item \textbf{Activity function:}
The activity function is a sequence of actions that are executed when the events associated with a trigger function occur. In the reference application, 
two action functions are defined. 
If the first trigger function is satisfied (i.e., $s^o_2$ is smaller than the threshold), then Action 1 is executed and the user's request is sent to the least loaded server, otherwise Action 2 is triggered and the request is forwarded to the controller. Both action functions are executed at the same switch where the request has been received. 
\end{itemize}

The actual implementation of the reference load-balancing application is shown in Listing~\ref{code:resource-lb-initial}. The listing highlights the simplicity of defining the application, by depicting how each of the previously discussed application elements can be defined and manipulated by the programmer thanks to the APIs provided by the LOADER programming model. 

\begin{table*}[!tb]
	\ra{1.5}
    \centering
	\caption{Example applications enabled by LOADER and their mapping the the LOADER programming model}
	\footnotesize
    \begin{tabularx}{\textwidth}{XXXXX}\toprule 
	Application & States & Reduction function & Trigger function & Action function  \\ \midrule\midrule
	DDoS detection \cite{netsoft} & Average rate of inbound SYN packets traversing each edge router & Sum of all states & Comparison against a fixed threshold & Controller notification \\ \midrule
	Distributed rate limiting \cite{raghavan2007cloud} & Average rate of inbound traffic traversing each edge router & Sum of all states & Comparison against a random threshold & Packet drop \\ \midrule
	Link-aware load balancing \cite{alizadeh2014conga} & Average load on uplink and downlink ports connecting a pair of two leaf switches & Argmin among the maximum of all uplink and downlink pairs sharing a common path & Change in the reduction function output & Insertion of a per-flow forwarding rule \\ \midrule
	Resource-aware load balancing \cite{lee2011load} & Instantaneous CPU utilization of servers & i) Argmin among all states, ii) Mean among all states & Comparison of the average global CPU utilization against a fixed threshold & Insertion of a per-flow forwarding rule if the average CPU utilization is below a threshold, controller notification otherwise. \\\toprule
    \end{tabularx}
    \label{tbl:other_loader_examples}
\end{table*}

Table~\ref{tbl:other_loader_examples} depicts other example applications which operate on replicated states. Those applications have been proposed in literature with either custom hardware implementation inside switches or by employing ad-hoc P4 code. We show how those applications can be easily mapped to the LOADER abstraction by employing the application element provided by our abstraction model. 
Detailed description of those applications, alongside with the evaluation of selected applications, will be presented in Sec.~\ref{sec:ddos}. 

For the sake of space we limit the discussion of the syntax and the implementation details of the proposed programming model, while focusing on its core functionalities. One core functionality is the explicit management of inconsistencies, which is specified as a parameter in both trigger functions (as explained in Sec.~\ref{sec:consistency_analysis}), and the semantic of the language which is discussed in the following.

\subsection{Semantics and order of execution}
By default, in LOADER all operations on states are executed in parallel. Such kind of dis-aggregation for the order of execution significantly reduces the embedding complexity as each element of the network application can be treated independently and will not require order synchronization.

Nevertheless, some application may require a specific order for the execution of the activity functions (e.g., appending new packet headers in a given order). Such kind of constraints may significantly increase the complexity of the overall approach by requiring additional ordering mechanisms whenever the activity functions are distributed across different switches. 

LOADER provides means of specifying particular order for the execution of operations by exposing to the programmer the \verb|SequentialActivityFunction| class. This class imposes hard constrains on the compiler forcing it to treat the enclosed activity functions as a single sequential activity function. This in turns forces the embedder to perform co-located embedding of those activity functions, forcing them to be embedded in the same network devices, ultimately permitting sequential execution.

\subsection{Compilation phase}
Although in our experimental evaluation we implemented a minimal proof-of-concept compiler, implementing a network application compiler compatible with a broad variety of different network devices requires an immense effort and in-depth knowledge about each device architecture. For this reason, for the purpose of this work we discuss what the compiler must perform and how the proposed framework facilitates its operations. 

Network applications are compiled through the \emph{LOADER Compiler}, as shown in Fig.~\ref{fig:LOADER_framework}. 
The compiler takes as input the network capabilities in the form of available {\em basic primitives}, and the user-defined application in the form of LOADER application elements. 
The catalog of available primitives depends on the specific network devices operating in the network and is stored in the resource management module of the network controller and it is updated through the network management plane, e.g., at device installation time.
The application is then represented by the compiler in the form of a DAG (Directed Acyclic Graph) composed of its basic elements, as shown in Fig~\ref{fig:LOADER_decomposition}. 
The compiler then reconstructs the dependency among each application element and maps them to basic primitives supported by the network devices composing the network so that, as depicted in Fig~\ref{fig:LOADER_decomposition}:
\begin{itemize}
	\item states are mapped into {\em  primitive data structures}, such as counters, registers, hash tables, etc., to store application states; % support
	\item reduction, trigger and activity functions are mapped into {\em primitive actions}, i.e. basic processing/decision capabilities offered by network devices. 
\end{itemize}

%For simplicity, unless explicitly specified, in the following we will refer to primitive data structures as states and primitive actions simply as actions.

\subsection{Optimal embedding and application reaction latency}\label{sec:l_embedding}
The embedding consists in mapping the primitive elements provided by the compiler into a set of physical network devices. This is performed by exploiting the target-specific drivers and southbound APIs (e.g., P4Runtime, gRPC, OpenFlow, etc.) offered by the embedding engine of the controller.

To perform the actual embedding, as depicted in Fig.~\ref{fig:LOADER_framework}, the embedder takes as input: i) the set of primitive elements provided by the compiler, ii) the resource availability inside the network provided by the controller resource manager and iii) the actual location of the resources inside the network provided by the controller topology manager. 
Given this information, it is possible to find a set of feasible embeddings of the decomposed application inside the network devices supporting the required primitives.
Notably, each element of the network application is not required to be embedded in a single network device. Instead, individual primitives composing the network application can be embedded in different network devices, based on the types of supported primitives, their amount and their location inside the network.
The adopted algorithm to optimize the embedding (i.e., computing the optimal number of replicas and their placement within the network) has been already investigated in our previous work~\cite{abu2019}, which serves as a natural integration to LOADER.
In the following we give insights regarding the functionalities and restrains of the embedding mechanism.

\subsubsection{Constraints on primitives location}
In the absence of co-location at the same network device of primitive actions and primitive data structures directly operated by those primitive actions, state replication is mandatory. Indeed, to perform the reduction of a given set of states, the states must be locally available at the network device operating the reduction function. This requires either to provide co-location of the states and reduction functions or to perform state replication at the network device storing the corresponding reduction primitive. 

\subsubsection{Inter-application state sharing}
\begin{figure}
	\centering
	\begin{subfigure}{.5\columnwidth}
		\centering
		\includegraphics[height=85pt]{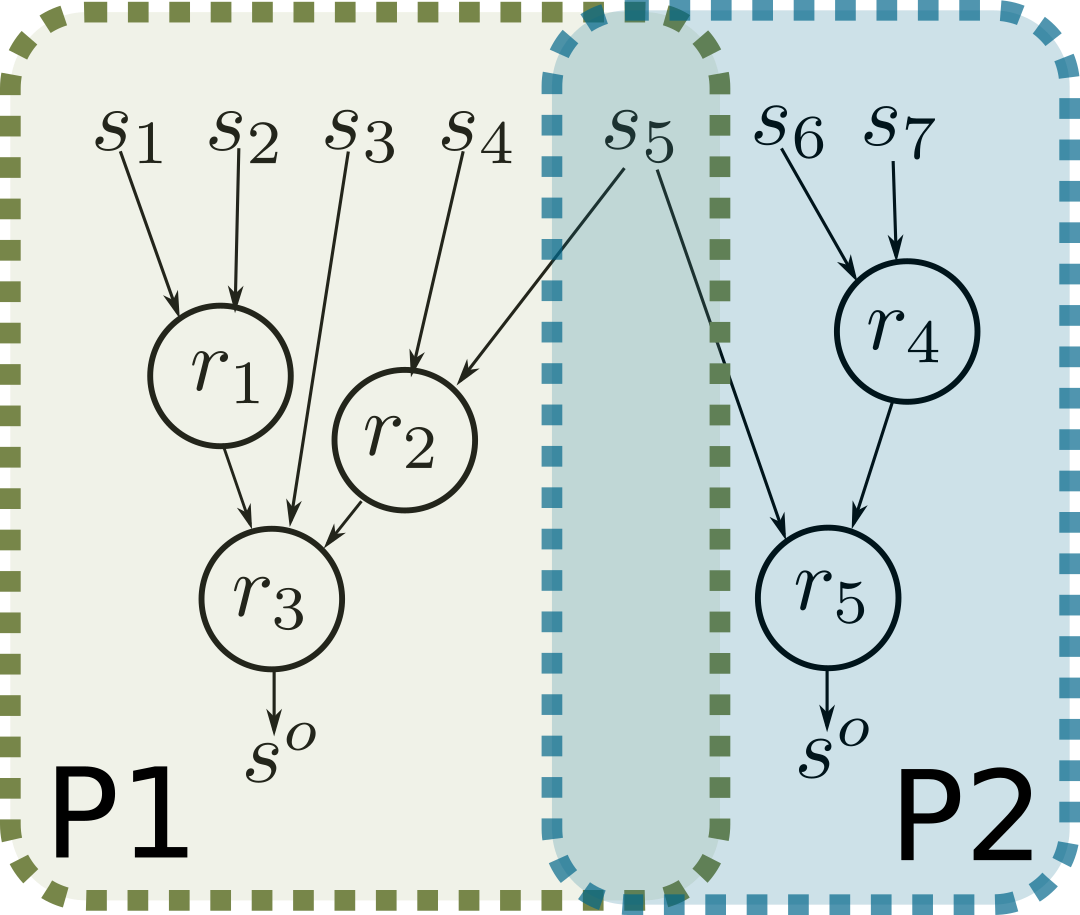}
	\end{subfigure}%
	\begin{subfigure}{.5\columnwidth}
		\centering
		\includegraphics[height=85pt]{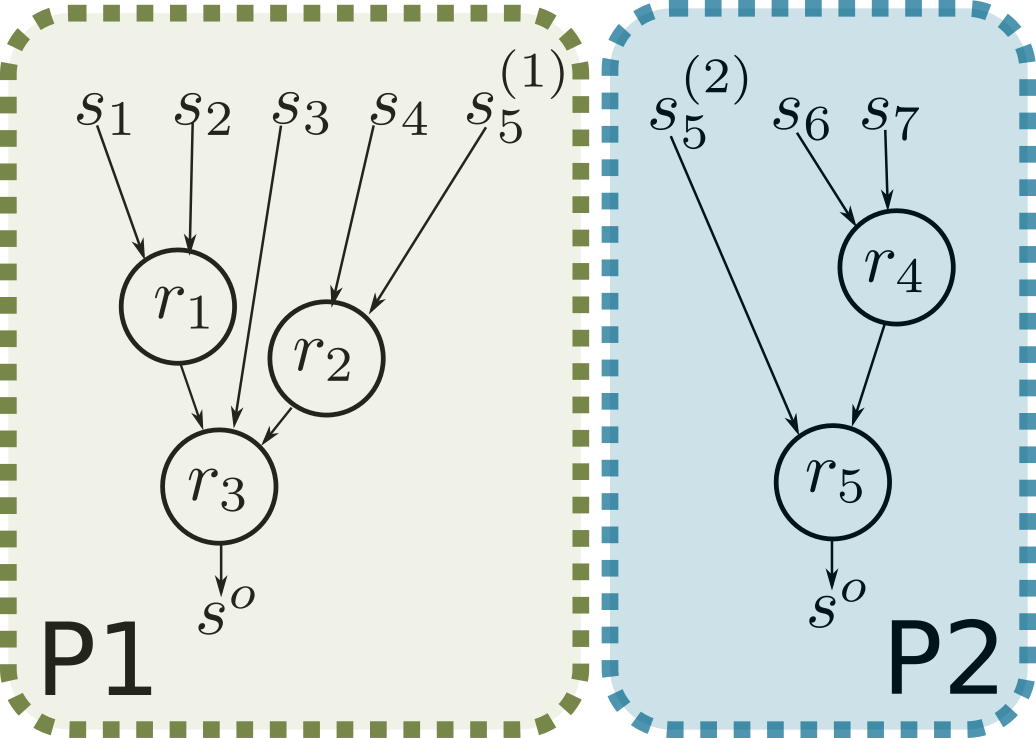}
	\end{subfigure}
	\caption{Reduction function decomposition in case of two network applications sharing a state $s_5$ without replicated states (left) and with replicated states (right).}
	\label{fig:reduction_decopmosition_mp}
\end{figure}

States may be shared among different network applications.
Fig.~\ref{fig:reduction_decopmosition_mp} shows an example of two network applications $P_1$ and $P_2$ sharing a common state $s_5$. Using a single replica approach, $s_5$ is required to be embedded into a single network device. As a consequence, the device storing $s_5$ must serve both $P_1$ and $P_2$, which, as previously discussed, may lead to scalability issues whenever the number of applications employing $s_5$ grows large. Instead, with state replication, the two applications can be made independent by replicating $s_5$ in $s^{(1)}_5$ and $s^{(2)}_5$. 
Note that the  concurrent access of two different applications to two replicas of the same state is equivalent to the concurrent access of two instances of the same application on such replicated states, as in the DDoS detection scheme discussed in Sec.~\ref{sec:ddos}.

\subsubsection{Application reaction latency}
Given an application embedding, it is possible to evaluate the corresponding {\em reaction latency}, by considering the position of the primitives in the network, the propagation delays between the involved network devices and the replication delay. For a single-replica state, the replication delay is by construction null as no replication occurs whatsoever. 
On the other hand, in the case of multiple replicas, the reaction latency models the latency required to propagate a new value of the state to all the replicas and will be explained in details in Sec.~\ref{sec:delay}. 
Interestingly enough, as investigated in~\cite{abu2019}, an optimal embedding might lead to multiple replicas. Although multiple replicas imply non-null replication delays, this delay can be compensated by a much smaller application execution latency. The distributed DDoS detection application, considered later in Sec.~\ref{sec:ddos}, is an example of such a scenario, clearly showing the advantage of keeping multiple replicas for some network-wide applications.

\subsubsection{Objective-based embeddings}
The optimal embedding is chosen by minimizing a particular cost function.
The definition of the cost function highly influences the way the embedding is performed, as shown by~\cite{abu2019}. 
As an example, a cost function aiming at reducing the network energy consumption or reducing the synchronization traffic between replicas may lead to scenarios in which the application is embedded into few network devices or eventually to a single network device (e.g., the SDN controller).
On the other hand, a cost function aiming at minimizing the network congestion may lead to multiple replicated states across different network devices to balance the traffic across the network. Thus, the definition of the cost function highly affects the level of distribution of the application, ranging from completely distributed implementations to completely centralized and stateless ones.

\subsection{LOADER in stateless SDN}
In the case of stateless SDN networks, with network devices able to perform only basic forwarding/routing operations, the LOADER approach is still viable. 
Indeed, LOADER provides only an abstraction layer between the actual application and its mapping to the network devices.
As previously discussed, the controller is seen as part of the available embedding targets during the embedding phase. Being typically a general purpose machine the controller is seen as a network device with unlimited computation resources, thus giving the embedder the possibility of eventually placing the network application at the controller. Nevertheless, as previously mentioned, the latency between the controller and the network devices is typically high as it includes both the network latency and the in-software processing delays at the controller. As shown in the following, this latency plays a fundamental role during the embedding as it directly affects the \emph{state inconsistency level} which is among the main user-defined constraints in LOADER.

\section{Bounding inconsistency among states}\label{sec:consistency_analysis}
To provide correct functionality of the application, all replicas of a state must be consistent. 
Consequently, a read operation of any replica at any given time should eventually return the same result.
The CAP theorem~\cite{cap} states that, for a replication scheme, out of Consistency, Availability and Partition tolerance, only two properties can be picked at the same time. Considering that network failures may occur, partition tolerance cannot be left out of the design of the replication algorithm, leaving us with two main reference models:
\begin{itemize}[leftmargin=*]
	\item {\em Strong consistency}.
	This model privileges consistency over availability, meaning that a read operation on any non-faulty replica will return the most recent committed value (same for all replicas) or an error. This property is achieved at the cost of reduced availability due to the requirement of multiple interactions between replicas and is based on complex consensus protocols~\cite{howard19}.
	\item {\em Eventual consistency}.
	This model privileges availability and results in instantaneous operations on all replicas with a considerably reduced protocol complexity.
	Although it introduces transient inconsistency, the latter can be seen as an error in the value of a local replica.
\end{itemize}

The choice between the two models depends on the level of tolerance of the considered network application in the presence of temporary inconsistencies between replicas of the same state.
{The majority of network applications require small packet processing latencies. Indeed, excessive latencies may lead to noticeable performance degradation in the case of real-time traffic and applications performing per-packet processing. This leads to the necessity of privileging high availability when state changes occur.}

For highly mutable states, replication schemes based on strong consistency may lead to excessive latency due to the complex protocol needed to reach the consensus, ultimately leading to excessive commit delays which will preclude the correct functionality of applications.
However, the majority of network applications operate on statistical network measurements and remain robust even in the presence of small errors for the value of the global state, making strong consistency less essential.

\subsection{Replication delays and state inconsistency}\label{sec:delay}
LOADER does not impose any constraint on the adopted replication scheme, leaving to the programmer the freedom of implementing any replication protocol alongside with the suitable reconciliation scheme supported by network devices. 
It is generally true that replication schemes based on strong consistency are more complex and introduce larger latency to commit a value than the schemes based on eventual consistency.
Thus, without loss of generality, in the following discussion, we focus on supporting eventual consistency. More in details we focus on the case of optimistic replication realized with basic gossiping for which the precise sequence of concurrent writes on the different replicas is not affecting the application correctness.

In an eventual consistency scheme, each state is associated with a certain {\em replication delay} $d_i$, i.e., the maximum amount of time required to convey a state update to all of its replicas. Note that $d_i$ corresponds also to the worst case inconsistency time. Assume now that a state is replicated with period $d^{R}_{i}$ (i.e., the inverse of the replication frequency).
Let $d^P_{nm}$ be the communication latency between network devices $n$ and $m$, taking into account the propagation delay (we assume isolation of replication traffic from data traffic, thus negligible queueing delays).  
If $\mathcal N_i$ is defined as the set of nodes storing replicas of $s_i$, we can claim:
\begin{equation}\label{eq:di}
d_i=d_i^R+\max_{n,m\in\mathcal N_i} d^P_{nm}
\end{equation}

The programmer is required to develop network applications by keeping in mind that different state replicas may suffer of inconsistency intervals during which their values may differ. To cope up with this, LOADER exposes to the programmer the possibility of defining an explicit inconsistency level for the replicated states. This is made possible by defining a level of {\em state inconsistency} inside the trigger function. The output of a  network application is driven by the outcome of the trigger function, and for this reason, specifying the inconsistency level at the trigger function is sufficient to determine also the overall state inconsistency of the application. 

We foresee two main inconsistency metrics which can be defined by the programmer: (1) {\em  time obsolescence} $\epsilon_t$ and (2) {\em update error} $\epsilon_r$. The former metric provides means of defining an upper bound on the time freshness of the state replicas and guarantees that at any given time any replica will contain a value not older than $\epsilon_t$ in time. The latter instead specifies the maximum admissible inconsistency in terms of uncommitted writes for any state variable, thus ensuring that the difference between all the replicated states does not exceed a number $\epsilon_r$ of state writes. 
The actual choice of the adopted inconsistency metric and the corresponding value is left to the programmer and it largely depends on the particular network application.

LOADER guarantees that the constraints specified by the programmer in terms of inconsistency metrics are satisfied.
%\todo{how would the developer decide on Er? How sensitive would the application be on this error (also in Section V.A, par. 4)?}
%by acting on one the two terms of~\eqref{eq:di}. 
During the embedding phase LOADER first assigns replicas positions in the network so to minimize the maximum communication latency %$d_{nm}^P$ 
between any pair of replicas, i.e., minimize the second term of~\eqref{eq:di}. Following this operation, two scenarios are possible.
If a time obsolescence $\epsilon_t$ is specified, then the replication periodicity $d_i^R$ must be set such that:
\begin{equation}\label{eq:d2}
d_i^R\leq \epsilon_t-\max_{n,m\in \mathcal N_i} d_{nm}^P
\end{equation}
If instead an update error $\epsilon_r$ is specified, now $d_i^R$ must be related to rate of write operations on the state over the time. To satisfy this constraint for a generic state $x$, it is sufficient to evaluate $\delta^*_\tau$ as the maximum number of write operations performed on $x$ over a time interval $\tau$. Note that  $\delta^*_\tau$ depends on the specific meaning of the considered state and should be evaluated a priori. E.g., for a packet counter at an interface, it is obtained by the data rate divided by the transmission time of a minimum size packet. 
Let $|x|_t$ denote the number of writes for state $x$ up to time $t$. By construction, it holds: % we can claim:
\begin{equation}\label{eq:2}
|x|_{t+\epsilon_t}-|x|_{t}\leq \delta^*_{\epsilon_t} \epsilon_t
\end{equation}  
By definition,  we can bound~\eqref{eq:2} with $\epsilon_r$ and obtain: 
\begin{equation}
\delta^*_{\epsilon_t} \epsilon_t\leq \epsilon_r
\end{equation}
Based on~\eqref{eq:d2}, $d_i^R$ is chosen such that:
\begin{equation}\label{eq:d5}
d_i^R\leq \dfrac{\epsilon_r}{\delta^*_{\epsilon_t}}-\max_{n,m\in \mathcal N_i} d_{nm}^P
\end{equation}

Note that in the case of states which permit a definition of absolute state error  based on some norm (e.g., scalars, arrays, graphs), knowing the nature of write operations permits to translate the update error into \textit{absolute value error}. Assuming that a write operation can alter the state by a maximum amount,  it is possible to rewrite $\delta^*_\tau$ in terms of absolute state variation and derive the temporal constraints following the same above formulation.

%After the embedding, LOADER tunes the remaining parameter of~\eqref{eq:di} by adjusting the 
%In the case of zero or unfeasible inconsistency levels, LOADER will employ only a single replica of a state.
%
%PG: We can claim $d_i<\epsilon_t$ and thus obtain $d_i^R$. 

Listings \ref{code:2ms_trigger}, \ref{code:2p_trigger} and \ref{code:no_trigger} provide an example of the definition of a trigger function in LOADER for a simple scenario (i.e., sum of two states). The listings show respectively a trigger function with a given value of time obsolescence ($\epsilon_t$), a trigger function with update error ($\epsilon_r$) and a trigger functions which does not tolerate any state inconsistency. 

\vspace{-5pt}\begin{center}\begin{minipage}[tb!]{0.9\linewidth}
\begin{lstlisting}[language=Python2, caption=Example of trigger function with time obsolescence $\epsilon_t$ equal to 2ms., label=code:2ms_trigger]
r = ReductionFunction(states=[s1, s2], operation=stateSum)
	
tr = TriggerFunction(s0=r.Result(), trigger=(r.Result() > 0), 
					inconsistencyLevel=TimeObsolescence(2, "ms"))
\end{lstlisting}\squeezeup
\end{minipage}\end{center}

\vspace{-5pt}\begin{center}\begin{minipage}[tb!]{0.9\linewidth}
\begin{lstlisting}[language=Python2, caption=Example of trigger function with update error $\epsilon_r$ equal to 10 writes., label=code:2p_trigger]
r = ReductionFunction(states=[s1, s2], operation=stateSum)
	
tr = TriggerFunction(s0=r.Result(), trigger=(r.Result() > 0),
					inconsistencyLevel= UpdateError(10))
\end{lstlisting}\squeezeup
\end{minipage}\end{center}

\vspace{-5pt}\begin{center}\begin{minipage}[!tb]{0.9\linewidth}
\begin{lstlisting}[language=Python2, caption=Example of trigger function without inconsistency (i.e. replication is not permitted)., label=code:no_trigger]
r = ReductionFunction(states=[s1, s2], operation=stateSum)
	
tr = TriggerFunction(s0=r.Result(),trigger=(r.Result() > 0))
\end{lstlisting}\squeezeup
\end{minipage}\end{center}

\subsection{Replication traffic generation}\label{sec:syn_traffic}
To replicate a state, network devices generate by themselves update packets, based on the required replication periodicity $d_i^R$. This generation is not currently supported in off-the-shelf hardware for stateful switches as a fundamental primitive, since, for performance reasons, packet generation events are triggered only by packet arrivals. Depending on the actual hardware, we foresee different solutions which provide a way of generating new packets without any hardware modification of current off-the-shelf chipsets, which are briefly discussed in the following.

\subsubsection{Controller-triggered updates}
The generation is triggered by the controller. In the case of periodic updates, the controller sends periodic trigger messages to the network devices, where they are processed and used to generate the update packets, by acting upon the reception of the trigger messages. 
Despite its simplicity, this approach has many limitations.
First, the required control bandwidth from the controller to each switch can become relevant for small update periods. Second, the controller is loaded with an additional task, impairing its scalability.

\subsubsection{Traffic-triggered updates}\label{implementation:triggering:event_driven}
The  generation is triggered directly by the reception of data packets received at any interface of the network device. This permits to self-adapt the amount of replication traffic on the dynamicity  of the states, whenever these depends on the arrived traffic.
In terms of implementation, the update message is  generated by cloning a data packet and then modifying it to carry the update value.
For what concerns stateful SDN switches, we consider two possible approaches to regulate the replication traffic rate based on native internal primitives:
\begin{itemize}[leftmargin=*]
	\item packet period $p$. By keeping a packet counter, a new update packet is generated every $p$ received packets, i.e., $d_i^R\leq p/r_{\min}$ where $r_{\min}$ is the minimum packet arrival rate over the whole switch. This can be used in~\eqref{eq:di} to choose $p$ and satisfy the given inconsistency metrics.  Intuitively, the update rate is proportional to the arrival rate of data packets which may suit well particular traffic-monitoring applications. On the other hand, for other applications this approach may lead to shortcomings, since in the absence of transit traffic no updates will be generated. 
	\item time period $\tau^R$. An update packet is generated at the first packet arrival after $\tau^R$ time and thus $d_i^R\leq \tau^R+1/r_{\min}$.  This can be used in~\eqref{eq:di} to choose $\tau^R$ and satisfy the given inconsistency metrics. Intuitively, this case results in periodic updates, i.e., a fixed replication rate approximately  independent from the traffic.
\end{itemize}

In terms of message format, the replication packet must carry the state identifier, the state value and the identifier of the switch originating the update. All identifiers can be predetermined by the controller at the time of application instantiation. This mechanism guarantees the state uniqueness while providing flexibility in term of state format encoding. 
Finally, to route properly the replication traffic, the position of each application primitive in the network is considered. LOADER exploits the network knowledge at the controller to install updates forwarding rules through a Steiner tree, either shared across all the states or one specific for each state.

%% file: lodge_implementation.tex
%!TEX root = main.tex
\section{LOADER implementation}\label{sec:LOADER_implementation}
%Since the scope of this work is to provide an abstraction for applications embedding we limited our implementation to a simple feasibility study. 
To prove LOADER feasibility, we developed a lightweight implementation of the framework. We integrated LOADER into ONOS v1.14 while using P4~\cite{bosshart2014p4} and Open Packet Processor (OPP)~\cite{OPP} switches for the data plane. 
The choice of these two distinct data plane architectures aims at showing the generality of the proposed approach, which results to be independent of the specific type of devices adopted in the network.

\subsection{Control plane implementation}
LOADER has been integrated inside the ONOS controller in the form of an ONOS application with custom control logic overriding the default controller behavior.

\subsubsection{Application definition}
We consider a set of predefined application elements supported by the switches.
This assumption permits to drastically simplify the implementation of the application definition phase inside ONOS. In particular, we specify each application element by means of predefined ad-hoc classes for each type of application element, based on the primitives supported by the switches. Thus no interaction with the resource manager of ONOS is performed.

\subsubsection{Application elements embedding}\label{sec:embedding_stainer}
For the purpose of this work, we consider a homogeneous network with  devices composed of programmable switches having the same type and amount of resources. 
Since the algorithm to solve the optimal embedding problem is out of the scope of this work, we consider the following simple embedding scheme, inspired to the one proposed in~\cite{abu2019}.
The position of each replicated primitive inside the network is determined by considering the betweenness centrality of each network device, weighted by the amount of traffic flowing through it. The main idea is to privilege the devices that are traversed by most of the traffic. 
Furthermore, the number of replicas of each primitive is fixed a-priori and not optimally chosen.
The replication traffic between the different replicas is routed on a single Steiner tree shared across all the replicas. This permits to reduce both the amount of replication traffic and the amount of flow table entries.

\subsubsection{State identification}\label{sec:identifiers}
%As motivated in the previous two sections, 
LOADER requires a unique identifier for each state, to guarantee correct processing of update packets.
Similarly to other network programming frameworks~\cite{yeganeh2016beehive} LOADER assigns a unique identifier to each state during the application compilation phase. For replicated states an additional identifier is assigned to distinguish between different replicas of the same state.

\subsection{P4 implementation}\label{sec:p4}
P4~\cite{bosshart2013forwarding}  is a novel data plane programming language which aims to achieve both target and protocol independence, in-field reprogrammability while providing also stateful operations thanks to the presence of persistent memories.
Similarly to OpenFlow, P4-enabled switches exploit a reconfigurable match-action pipeline, thus permitting to define multiple packet processing stages.
% Consequently the language specifications are designed to permit fast and efficient translation of the programmer-defined features in match-action rules. %, to guarantee low computational overhead and line-rate processing speed. 
P4 is protocol independent thanks to the presence of a programmable parser and deparser placed at the two extremes of the packet processing pipeline. Thanks to the parser programmability, it is possible to define custom protocol headers or even extend the parsing/deparsing actions to the packet payload.

To provide connectivity between ONOS and  P4 switches (version~1.1), we exploited P4Runtime.
At the time of this work, P4Runtime implementation in ONOS~v1.14 performs only basic flow tables manipulations without providing support for features such as runtime pipeline modification and manipulation of extern objects such as registers and counters. Due to these limitations,  we implemented the required primitive data structures and the replication control logic directly in P4 
%PG: implement the LOADER logic
instead of letting the controller push them to each switch at application creation time. However, the controller is left with the possibility of activating or deactivating application elements inside a switch, which is equivalent to pushing new logic.

\subsubsection{Replication traffic format}
%As mentioned in Sec.~\ref{sec:identifiers}, to provide uniqueness and absence of ambiguity for replication traffic, the employed packet header format requires a set of identifiers. 
Replication traffic is transported through packets that are formatted with a custom header carried by Ethernet packets, identified by an unused protocol type (\verb|LOADER_ETHTYPE|) in the Ethernet header.
We leverage P4 to define custom packet formats and we implemented LOADER header format directly inside the programmable parser.
%LOADER header is placed between Ethernet and IP headers and the Ethernet type field is set to a currently unused one to permit unequivocal identification of LOADER packets.

Listing~\ref{code:LOADER_header_p4} shows the full header format of LOADER packets. As previously mentioned, all identifiers are assigned by the controller during application initialization. Being \verb|srcSwID|, \verb|stateID| and \verb|replicaID|, respectively, source switch, state, and replica identifiers, which are required to correctly interpret and process the update packets at the destination switches. 
On the other hand, the inclusion of \verb|dstSwID| permits to implement more sophisticated replication schemes instead of employing ours based on shared spanning trees. In our experiments we implemented a broadcast transmission among all switches holding the replicas and for this reason \verb|dstSwID| field remained not utilized. 
The \verb|stateValue| field carries the actual value of the replicated state and its length is upper bounded by a constant number of bit, i.e., \verb|STATE_MAX_WIDTH|.
Finally, the \verb|L3ProtocolType| field permits to attach LOADER packets to transit packets, i.e., to piggyback replication information on data traffic. % This provides the possibility of appending state-update packets on top of transit packets and 

We generate nested LOADER headers to carry multiple state updates in a single packet. 
This functionality is depicted in Listing~\ref{code:LOADER_parser_state} which shows the implementation of the LOADER protocol parser. %After extracting LOADER information, depending of the value of \verb|nextHeaderType|, it is possible to extract subsequent LOADER headers. At the same time it permits to extract the original encapsulated packet (if any). 
Although in this work we opted to define a custom LOADER header, replication traffic transport can be also implemented by employing Inband Network Telemetry~(INT) format~\cite{kim2015band} defined by the P4 Language consortium.

\begin{figure}[!tb]
%\vspace{-5pt}\begin{center}\begin{minipage}{0.9\linewidth}
\begin{lstlisting}[linewidth=0.95\columnwidth,language=P4, caption=LOADER header definition in P4, label=code:LOADER_header_p4]
header LOADER_t {
	bit<32> srcSwID;
	bit<32> dstSwID;
	bit<32> stateID;
	bit<32> replicaID;
	bit<STATE_MAX_WIDTH> stateValue;
	bit<16> L3ProtocolType;
}\end{lstlisting}\squeezeup
%\end{minipage}\end{center}
\end{figure}

%\vspace{-5pt}\begin{center}\begin{minipage}{0.9\linewidth}
\begin{figure}[!tb]
\begin{center}
		\begin{lstlisting}[linewidth=0.95\columnwidth,language=P4, caption=LOADER parser implementation in P4, label=code:LOADER_parser_state]
state parse_LOADER {
	packet.extract(hdr.LOADER);
	transition select(hdr.LOADER.L3ProtocolType){
		LOADER_ETHTYPE : parse_LOADER;
		IP_ETHTYPE : parse_IP;
		default : accept;
	}
}\end{lstlisting}\squeezeup
%\end{minipage}
\end{center}
\end{figure}

\subsubsection{Generation of periodic update packets}\label{sec:synup}

Commercial implementations of stateful switches generally do not support the generation of self-triggered events, precluding the possibility of 
%neither arbitrary internal polling {\color{red}PG: polling???} nor the generation of new packets, thus precluding the possibility of 
employing periodic updates. 
%This makes the generation of periodic replication traffic challenging on off-the-shelf chips. 
However, in conformity with their purpose, switches are able to execute routines during packets reception and departure. Such routines may be related to simple packet processing up to more complicated user-defined routines in programmable switches. This behavior can be exploited to provide a simple mechanism to approximate a periodic traffic generation without hardware modifications.

%Due to the fact that programmable switches can execute arbitrary code when a packet is received on any of its ports it is possible to 

We exploit traffic-triggered updates, as described in Sec.~\ref{sec:syn_traffic}, in which the % transit traffic as a triggering mechanism for the generation of replication traffic, in which the 
% In typical scenarios a switch observes an aggregate incoming traffic with average rate $L>0$ pkts/s. Assuming constant $L$, this means that internally the switch executes a packet processing routing every $1/L$ seconds. A sufficiently large $L$ permits to implement fine-grained periodic user-defined routine execution without introducing any modification to the underlying hardware. 
%
temporal periodicity $d_i^R$ is obtained as follows.
%Given $L$  and the desired replication periodicity $d^R$ it is sufficient to employ a counter which will trigger the periodic replication routine once $n=\lceil d^R\times  L\rceil$ packets have been observed since the last replication routine execution.
%It is true that in real case scenarios $L$ is not constant and it is instead variable in time. For this reason a naive model employing a simple packet counter is not enough to guarantee correct functionality of the mechanism. To overcome this issue we implemented a solution which employs time registers in conjunction with the internal switch clock. 
During the execution of a replication routine, the current timestamp $t_\text{clk}$ is saved as $t'$. %\todo{Revise notation}
For each subsequent incoming packet we check the value of the internal clock $t_\text{clk}$ and compare it against the expected execution time of the routine, i.e., against $t'+d_i^R$. If $t_\text{clk} \ge t'+d_i^R$ a new replication routine is executed generating an update packet and $t'$ is updated.
Consequently, the first packet arriving after $d^R_i$ time will trigger the generation of the update packet.

%\subsubsection{Generation of a replication packet}
%In the case of replication traffic generation the periodic routine consists of 
The replication routine
generates and transmits a state-update packet filled with the state related information. To generate these packets we employ the packet-cloning extern provided in P4 v1 model~\cite{p4-repo}. Once the update has been triggered by an arriving packet, such packet is cloned to the egress port that has been assigned to it by its prior processing. Subsequently the original packet undergoes a transformation which substitutes its original header with the LOADER header filled with all the information related to the state which needs to be updated. At the same time the payload of the packet that triggered the update is dropped. Following this operation, the newly created LOADER packet is transferred to the corresponding output queue without undergoing further processing. Since the triggering packet needs to be fully processed at the time of cloning, this functionality, which is illustrated in Listing~\ref{code:LOADER_cloning}, resides at the very end of the ingress processing pipeline. In this way the replication traffic generation routine does not impact in any way the transit packets.

\subsubsection{Replication traffic routing}\label{sec:syn}
The generated replication packets are transmitted on one or more egress ports following a Steiner tree shared among all replicas. %The choice of the ports depends on the link belonging to the Steiner tree which is used to distribute all the updates across the replicas of a state and it is precomputed by the controller. 
The distribution tree consists of a mapping {\em (Switch, PortList)} which assigns to each switch of the Steiner tree the set of ports connected to the corresponding links. 
%over which they connect to other switches requiring the replication traffic. 
All newly generated or transit LOADER packets match against a specific match-action table which sends a copy of the packets for each port specified in  {\em PortList}. To avoid loops for transit LOADER packets,
% i.e. incoming LOADER packets which have not been generated locally, 
at the egress stage the original ingress port of each packet is compared against the current egress port. If the two ports are the same, the packet is dropped. This mechanism permits to keep the amount of flow entries related to LOADER routing as low as one entry per state per switch. %In the extreme case of a single distribution tree across all states, one entry is enough for all the replicated states.
\vspace{1em}

Both the P4 switch and the LOADER framework implementations are publicly available at the LOADER repository~\cite{loader-repo}. 

%\vspace{-5pt}\begin{center}\begin{minipage}{0.9\linewidth}
\begin{figure}[!tb]
\begin{lstlisting}[linewidth=0.95\columnwidth,language=P4, caption=Generation of replication packet in P4, label=code:LOADER_cloning]
if( meta.LOADER_meta.state == UPDATE_NEEDED ){
	clone_pkt_to_egress(sm.egress_spec);
	fillLOADERHeaderTable.apply(meta.LOADER_meta.state_id);
	set_state_update_time(meta.LOADER_meta.state_id);
}\end{lstlisting}\squeezeup
%\end{minipage}\end{center}
\end{figure}

\subsection{OPP implementation}
Open Packet Processor (OPP)~\cite{OPP} is a programmable data plane abstraction in which Extended Finite State Machines (EFSM) are used to model stateful forwarding algorithms. The OPP machine model extends the match-action tables pipeline model assumed by OpenFlow. Such tables are substituted with \emph{stages}, which can be either stateless or stateful. A stateless stage is in fact an OpenFlow-like match-action table. The pipeline processes packet headers to define corresponding forwarding behaviors. The packets are processed by the ingress pipeline, which is composed by a parser stage and several stateless and stateful blocks, after the processed packets go into the internal switch memory that holds the packet queues.

An OPP application requires the definition of the following components:
\begin{itemize}
    \item \textit{Lookup/update extractors}: these two blocks are configured by defining a combination of packet fields that are used to retrieve/update flow state information.
    \item \textit{Conditions}: conditions are arithmetic comparison operations of global/local variables and packet header fields; conditions are matched in the EFSM table along with the flow state and packet fields. 
    \item \textit{EFSM table}: programming the EFSM table requires the definitions of a set of EFSM entries formed by a \emph{match} section (as defined in the list item above)  and an \emph{action} section, which defines the state transition and a set of packet actions (drop, push header, forward, etc.) and update functions over the local registers. The EFSM table is configured as a standard OpenFlow table and is usually realized in ASIC switches using TCAMs.
    \item \textit{Global data variables}: OPP global variables are independent of a particular flow and can be used in the condition block.
\end{itemize}

The OPP protocol used between the OPP switches and controllers is a modified version of OpenFlow~1.3 standard, extended to support the configuration of an OPP pipeline. In particular, the configuration of the lookup/update extractors and the conditions are realized with two new experimental OpenFlow messages that carry the list of packets fields to be extracted from the packet headers and the arithmetic operations whose results are matched from the EFSM tables. Furthermore,  the configuration of the EFSM table requires the extension of the OpenFlow FLOW$\_$MOD message to support new match fields (conditions, and flow states) and new actions (state transition and data variables updates).

The OPP switch implementations is publicly available at the OPP source repository \cite{OPP-repo}. 

\subsubsection{Replication traffic format}
In the OPP prototype, we decided to format the replication packets by employing the 20 bit labels provided by the MPLS protocol. 
This design choice was taken for mainly two reasons: the MPLS header is a widely used protocol supported by most of the Internet nodes, and in our OPP implementation it was simple to handle such encapsulation header since adding a custom protocol would have resulted in a static implementation of the parser code to support a custom header. 
The Switch ID is encoded in the source and destination fields of the overlay IP protocol, assigned to each node by the control plane at configuration time. The State ID is inserted in the Experimenter field of MPLS (3 bits) and as such, confined to a maximum of 8 different states supported. Finally, the MPLS label (20 bits) carries the State Value. 

\subsubsection{Generation of periodic update packets}
OPP does not support a time-based generation of periodic events so, as in the P4 implementation discussed in Sec.~\ref{sec:synup}, the generation of time-related events is triggered only by the reception of packets.
To emulate a timer expiration we use per-flow registers to store the time difference between packet arrivals. This difference is then compared with  the replication period $d^R$. The result of this comparison is then matched by the EFSM table, resulting in the execution of the corresponding action present in the  table entry. % associated with that condition outcome.

To generate the replication traffic, we implemented two approaches.
In the first approach, we clone the arrived packet to generate an update packet. 
When the packet generation event is triggered, the cloned packet is attached with an MPLS header containing the correct state information, while the original packet continues its normal processing. 
In the second approach, the update packet is instead generated ex-novo by using a predefined template that already contains the MPLS header. The header fields are then modified according to the state information to be written in the packet. Differently from the first approach, this one has the advantage of reducing the size of update packets since they do not carry any data above the MPLS layer.

\subsubsection{Replication traffic routing}
As discussed in Sec.~\ref{sec:syn} for the P4 implementation, the mapping switch-to-output port to route the replication traffic is statically assigned by the controller at configuration time. In such a way, the forwarding decision is taken through the normal OpenFlow stateless match-action strategy.

%% file: mapping.tex
%!TEX root = main.tex
\section{Implementation and evaluation of network applications with LOADER}\label{sec:ddos}

As a proof of concept, we used the LOADER programming model to developed a simple yet significant application for the \emph{distributed detection} of  Distributed Denial of Service (DDoS) attacks, denoted as DDoSD. % (Distributed DDoS 
%
%On the Traditional implementations based on 
%The DDoSD application presents considerable implementation challenges related to reactivity and controller overhead in the case of traditional SDN and to data overhead in the case of solution based on single state replica.
%We validate the effectiveness of LOADER in a DDoSD use case.  
%Due the distributed nature of the attack, the 
%
%In the case of a DDoS attack, a large network (e.g., an autonomous system (AS) observes a sudden increase in the incoming rate of a particular type of traffic. 
%
The main idea of the distributed detection is to exploit the typical temporal correlation between the increase  of traffic across all the network devices at the border of the network, due to the distributed nature of the attack.
%In case of ASs with multiple edge routers the sudden traffic increase will be observed on multiple points at the same time. 
Clearly, the correlated traffic increase across the edge routers is a much more reliable way to detect an attack with respect to a monitoring the traffic on a single network device only. Consequently a network application performing DDoSD must be able to capture this sudden increase in the network traffic. 

With traditional SDN approaches, the controller is involved in the detection process by being notified about the transit packets by switches. This leads  to large overhead in terms of traffic and of detection latency. Instead, LOADER enables a {\em distributed} detection process operating directly at the switches, without any controller involvement. Furthermore, the actions to counter the attack are executed in a distributed way, by each network device involved in the detection. 

As shown in Fig.~\ref{f:topo}, we consider a large network (e.g., an Autonomous System - AS) connected to other networks (e.g., other ASs) through different edge routers and the attack targets a set of internal servers.
Since the definition of a realistic DDoSD algorithm is a well-known problem in the literature~\cite{zargar2013survey} and it is completely out of the scope of this work, we employ a simple proof-of-concept threshold-based detection scheme, which demonstrates the correct operation of the replication mechanism and can be used as a foundation for more sophisticated DDoSD algorithms. 

\subsection{Network application definition}\label{sec:ddos_application}
The total traffic entering the whole network and directed toward the targeted servers is defined as the sum of the inbound traffic over each edge router (\verb|SW1|-\verb|SW4| in our reference topology). Based on the value of the inbound traffic the network application must perform some retaliation to counteract the DDoS attack. Consequently it is straightforward to map this kind of application to a LOADER application as described in the following.
\begin{itemize}
	\item \textbf{States}: Given $N$ edge routers, we define $s_i$ as  the average rate of inbound traffic traversing the border router $i$, with $i=1,\dots,N$. As monitoring target, we employ the rate of incoming SYN packets directed towards the internal servers.
	\item \textbf{Reduction function}: The reduction function employed by the application is composed of a single primitive action, namely $r_1 = \text{sum}()$. Consequently, the output of the reduction function is defined as $s^o=\text{sum}(s_1,\ldots,s_N)$.
	\item  \textbf{Trigger function}: Following the previous discussion, we define the threshold function simply as a simple comparison of $s^o$ against a predefined threshold. Thus, a DDoS attack is detected locally at each switch if $s^o$ is larger than a given threshold, above which the attack is considered as detected. The threshold is determined with standard test-based statistical methods. 
	\item \textbf{Activity function}: We employ a simple activity function which notifies the controller once the application has been triggered.
\end{itemize}

The implementation of the DDoS application with LOADER programming model is available in~\ref{appendix:codes}.

\subsection{Benefit of replicated states}
{
In a single replica approach (i.e., in the absence of LOADER) the DDoSD application would require all the traffic entering the network  to traverse a single switch holding the state monitoring the incoming traffic. %Assuming that a single switch can process the required amount of traffic
Thus the network load would grow, increasing the congestion, and could not be compatible with some traffic management schemes (e.g., load balancing)  that require to control the routing arbitrary within the network. 
%still introduce significant overhead in terms of data traffic since all traffic will be required to traverse a single switch, thus forcing the routing within the same network. Consequently,  many essential traffic management schemes (e.g., load balancing) that control the routing across the network  cannot be applied to the will become inapplicable.

LOADER instead permits to replicate the entire DDoSD application over multiple switches, thus minimizing the data overhead over the whole network. At the same time, LOADER introduces an overhead in terms of replication traffic, whose amount depends on the allowed inconsistency level. The replication traffic will be evaluated experimentally for the DDoSD application in Sec.~\ref{sec:expe}.

Notably, DDoSD is robust to possible transient inconsistencies between the values of total traffic estimated at each switch, thus employing an eventual consistency replication scheme will not create noticeable degradation due to replicated states estimation errors.

\subsection{Implementation}
The considered DDoSD scheme has been implemented on top of two different programmable data plane platforms: (1) $P4_{14}$/$P4_{16}$; (2) OPP.
Furthermore, the definition of the DDoSD application was performed inside ONOS with LOADER abstraction which permits to automatically offload and configure the developed network application.

\subsubsection{Control plane implementation}
We implemented basic LOADER functionalities related to this particular use case inside ONOS. 
We employ the routing algorithms and the embedding mechanism based on betweenness centrality discussed in Sec.~\ref{sec:embedding_stainer} with a maximum amount of admissible replicated states equal to $C$.
We assume a sufficiently large amount of resources inside switches, thus permitting function co-location with consequent replication of all application elements.

\subsubsection{Data plane implementation with P4}\label{sec:ddos_p4}
Our prototype is developed and tested in a virtual environment using Mininet~\cite{lantz2010network} and P4-enabled virtual switches targeting using the v1 Model and using the Simple Switch Architecture~\cite{p4-repo}. 
We estimate the rate of incoming TCP SYN packets by employing a sampling window equal to $\delta$. Let $r_k(t_n)$ be the estimated rate in the time interval $(t_n - (k+1)\delta, t_n - k\delta]$ with $t_n = n\delta, n\in\mathbb{N}$. The average rate is estimated at each switch $i$ as
%\begin{equation}\label{eq:exponential_average_bits}
%s_i(t)=\dfrac{1}{w}\sum_{j=0}^{w-1} r_{n-j},\qquad \text{for }t\in[n\delta_r,(n+1)\delta_r) 
%\end{equation}
\begin{equation}\label{eq:exponential_average_bits}
s_i(t_n)=\dfrac{1}{w}\sum_{k=0}^{w-1} r_k(t_{n})
\end{equation}
and represents the local state to be shared across all the other border routers, coherently with  the description of Sec.~\ref{sec:ddos_application}.
In particular, $w$ is chosen as a power of 2 due to the hardware limits in P4 switches imposed to the types of operations that can be implemented, i.e., shift operations are supported, divisions are not~\cite{sharmap4}.
%easily implementable with a basic shift operation, supported by P4. 
Notably, 
The $w$ most recent samples of the estimated rate are stored in a circular buffer. 
Replicated states are instead saved in dedicated registers.

\subsubsection{Data plane implementation with OPP}\label{sec:opp}
The OPP implementation requires a sequence of three stages: stage 0 extracts the state from update messages; %and detects flagged packets (to avoid double counting); 
stage 1 stores the state from the metadata notified by the previous table, performs monitoring and detection and generates update messages; stage 2 performs simple L3-forwarding.
Stage 0 represents the stateful processing core of replicated states. The processed flows are identified by the IPv4 destination addresses of the target servers. Stage 0 also considers one flow data variable containing the switch-local state and the $C-1$ variables storing the replicated states. Switch-local state $s_i$ is computed by employing a hardware-implemented Exponential Weighted Moving Average (EWMA) counting the number of TCP SYN packets in a given preconfigured time window.
%The application is executed whenever a packet belonging to the replication traffic is received carrying updates of the remote replicas.

\subsection{Experimental evaluation and validation}\label{sec:expe}
We configure a Mininet-based emulation environment deploying the topology shown in Fig.~\ref{f:topo}, where, for the sake of simplicity,  each cluster and each AS is represented by a Mininet host. 
%We run the DDoS detection scheme  described in Sec.~\ref{use_case:implementation}. 
To simulate the DDoS attack,  we use \texttt{hping3} tool to send TCP SYN requests from all ASs to all internal servers. In each experiment, during the first 20 seconds, we send the request at a slow rate, and then we increase the rate of all senders in a such a way to trigger the execution of the activity function. 
We consider experiments with varying $C$: (i) single replica embedded in SW1 ($C=1$), (ii) 2 replicas ($C=2$) embedded in SW1 and SW3, and (iii) 4 replicas ($C=4$) embedded in SW1, SW2, SW3, SW4. We repeated the experiments to achieve negligible 95\% confidence intervals if shown in the plots.

\begin{figure}[tb!]
	\centering
	\includegraphics[width=.8\columnwidth]{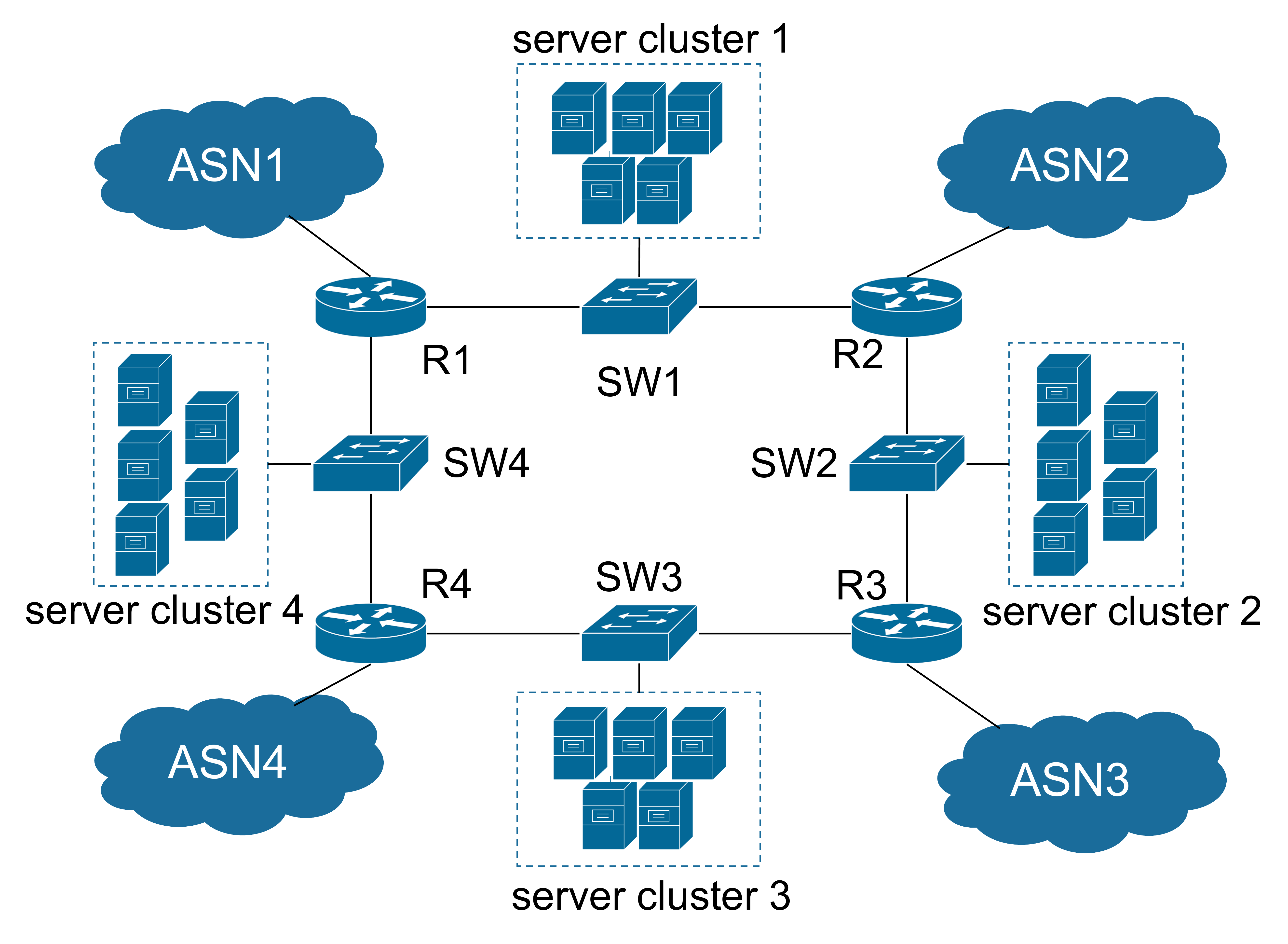}
	\caption{Reference topology for DDoS Detection use case.}
	\label{f:topo}
\end{figure}

\begin{figure}[tb!]
	\centering
	\includegraphics[width=.99\columnwidth]{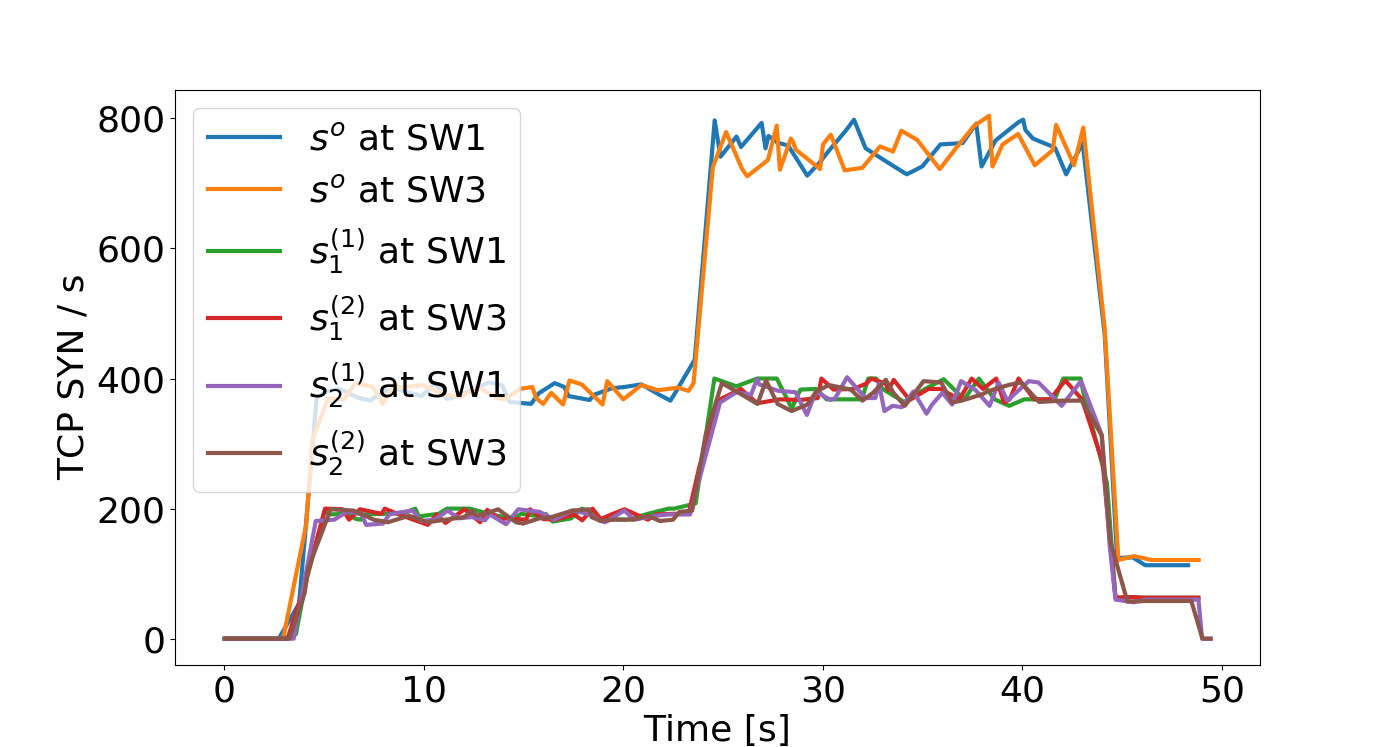}
	\caption{Temporal evolution of the local, remote and global states for the stateful switches in case of 2 replicas for the global state in P4 implementation.}
	\label{fig_states}
\end{figure}

Fig.~\ref{fig_states} shows the evolution of application states $s_i$ alongside with the evolution of $s^o$ for the case of 2 replicas, implemented in P4. Identical results are obtained with OPP and thus are not reported for the sake of space. As expected, the values of $s^o$ evaluated at SW1 and SW3 are coherent, and permit a contemporary detection of the DDoS attack in the two switches, without any interaction with the controller. This experimental result validates our proposed implementation for both P4 and OPP. 

In Figs.~\ref{fig_links_P4}-\ref{fig_links_OPP} we show the average utilization of the links present in the ring topology connecting all switches, for different values of $C$, with both P4 and OPP implementations. 
Clearly, for one replica (i.e, single replica approach) the load on the link is greatly unbalanced and in general higher for all the links. 
%The different behavior between P4 and OPP for some links (e.g.,\ SW1-R2) are due to the different routing schemes adopted in the topology.
By increasing the number of replicas to $2$, the load of the data traffic decreases by a factor of $1.6$ both in P4 and OPP and is much better balanced across the links. The slightly different values depends on the different mechanisms adopted for triggering the update event by the incoming traffic: in P4 the update rate depends on the traffic, whereas in OPP it is independent.
%traffic triggered, %generated proportionally to the traffic reaching the stateful switches, 
%whereas in OPP they are periodically generated at constant rate. 
Adding two other replicas reduces the data traffic by around 20\% in both implementations, but now the replication traffic becomes more relevant due to the higher number of replicas. Indeed, the fraction of update packets increases from 14\% (for 2 replicas) to 24\% (for 4 replicas) in P4 and from 11\% (for 2 replicas)  to 23\% (for 4 replicas) in OPP. 
%Notably, in OPP the \syn traffic doubles from 2 to 4 replicas equally for all the links, thanks to the periodic update triggering, whereas in P4 the growth in the \syn traffic depends on the actual load on the stateful switches.  
Thus, the two implementations behave very similarly and show the beneficial effect on the overall traffic in the network due to multiple replicas.

\subsection{LOADER-induced overhead}
As previously discussed and shown in Fig.~\ref{fig_links_P4} and  Fig.~\ref{fig_links_OPP}, LOADER adds some network overhead in the form of added synchronization traffic. The actual characterization of the amount of synchronization traffic highly depends on the network topology and the definition of the state. The impact of those factors has been exhaustively analyzed in our previous work~\cite{abu2019}.

From the point of view of device resource utilization, the amount of memory required to manage replicated states scales linearly with the degree of replication. Specifically, every switch must store its own local state values and the remote state values. Alongside those states, switches must also store an aggregate value combining local and remote states into a global state, which is then used as input to the reduction function. This translates into a total requirement of $A(C+1)$ bits of register memory per switch, being $A$ the size in terms of bits of a generic state to replicate (e.g.,  $A=32$ bits in the case of simple counters considered in the DDoS use case).

\begin{figure}[tb!]
	\centering
	\includegraphics[width=0.85\columnwidth]{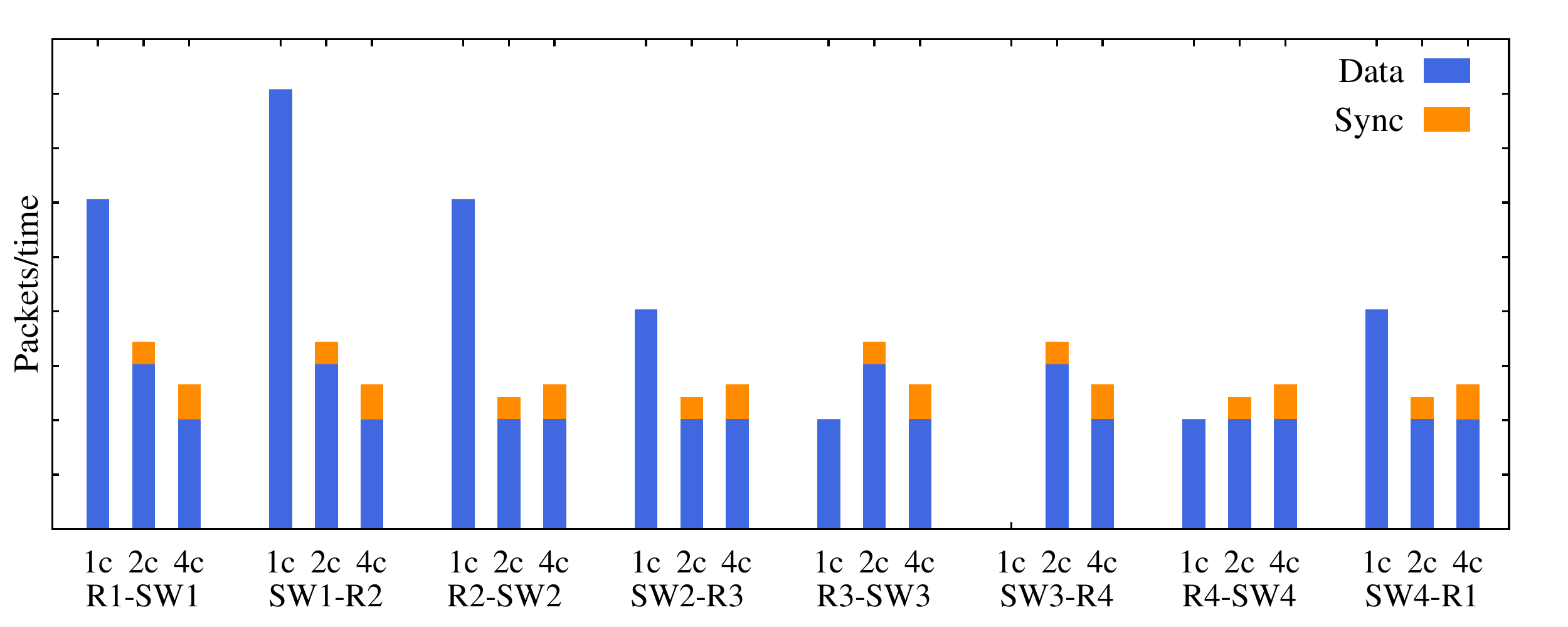}
	\caption{Average link utilization for data and for replication traffic in case of $1, 2, 4$ replicas for global state in P4 implementation.}
	\label{fig_links_P4}
\end{figure}

\begin{figure}[tb!]
	\centering
	\includegraphics[width=0.85\columnwidth]{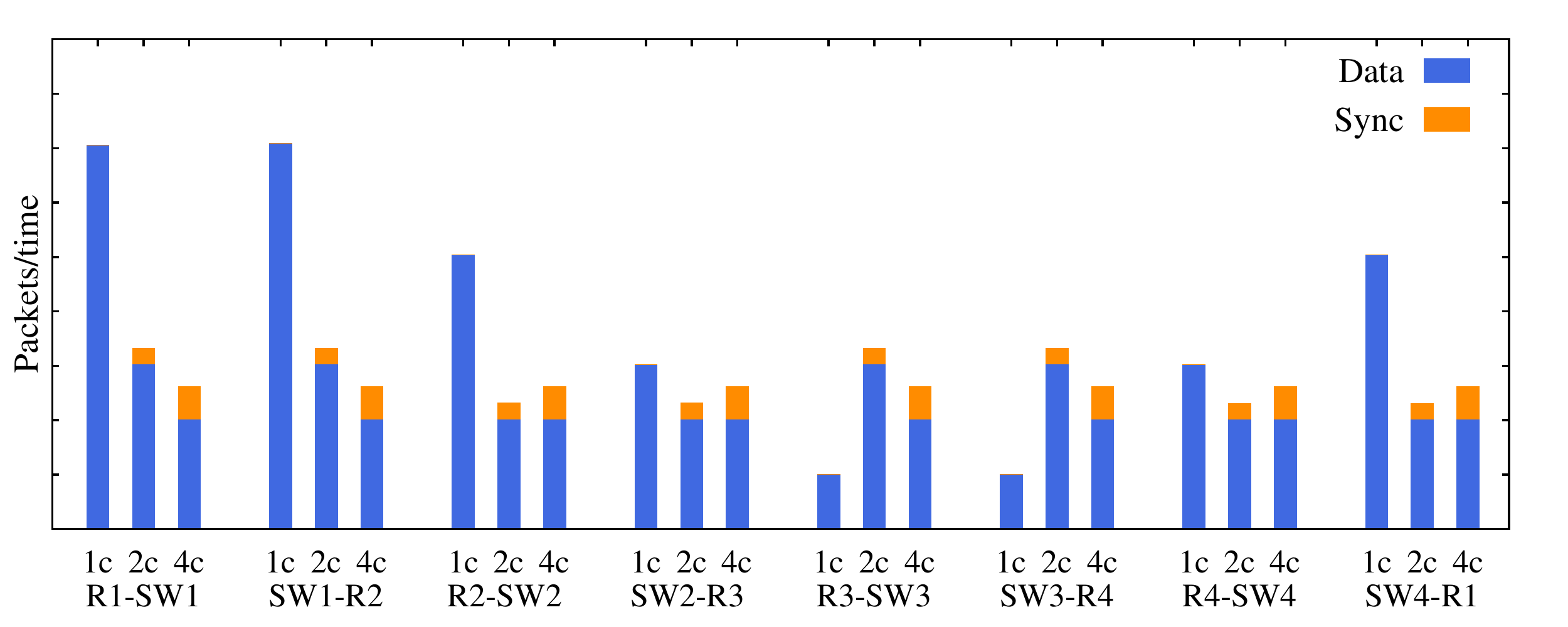}
	\caption{Average link utilization for data and replication traffic in case of $1, 2, 4$ replicas for global state in OPP implementation.}
	\label{fig_links_OPP}
\end{figure}

%% file: other_applications.tex
%!TEX root = main.tex
\subsection{Other applications enabled by LOADER}\label{sec:other_apps}
Although being significant, the DDoS use-case does not highlight the whole versatility of the proposed programming models. For this reason, in the following we describe some examples of network applications (previously described in Table~\ref{tbl:other_loader_examples}) which are shown to benefit from state replication. We show how those applications can be implemented with LOADER by providing their elements mapping and a code example for each of them. The actual implementation of those applications with the LOADER programming model is presented in~\ref{appendix:codes}.

\subsubsection{Distributed rate limiting}
In~\cite{raghavan2007cloud} the authors propose a network-wide global token bucket. Similarly to a local token bucket, a global one permits to rate limit all the incoming traffic in a given network thanks to a network application performing probabilistic dropping at the edge routers of the network. However, differently from a local one, a global token bucket involves an instance of the same token bucket run independently at each border router and using a single shared state accounting for the {\em total} inbound traffic.

This kind of application can be easily mapped to LOADER by considering the DDoSD scheme and by changing only the trigger and the activity functions as follows:

\begin{itemize}
	\item \textbf{States}: Given $N$ edge routers, we define state $s_i$ as the average rate of inbound traffic traversing edge router $i$, with $i=1,\ldots,N$.
	\item \textbf{Reduction function}: The reduction function performs a sum operation among all local state $s_i$ with $s^o=\text{sum}(s_1,\ldots,s_N)$.
	\item \textbf{Trigger function}: In order to perform probabilistic dropping the trigger function must invoke the activity function proportionally to the rate of the incoming traffic and the desired rate.
	\item \textbf{Activity function}: Identically to the DDoSD case, the activity function must perform dropping of incoming packets whenever invoked as to guarantee that the total incoming traffic is less than a given threshold.
\end{itemize}

\begin{figure}[tb!]
	\centering
	\includegraphics[width=0.95\columnwidth]{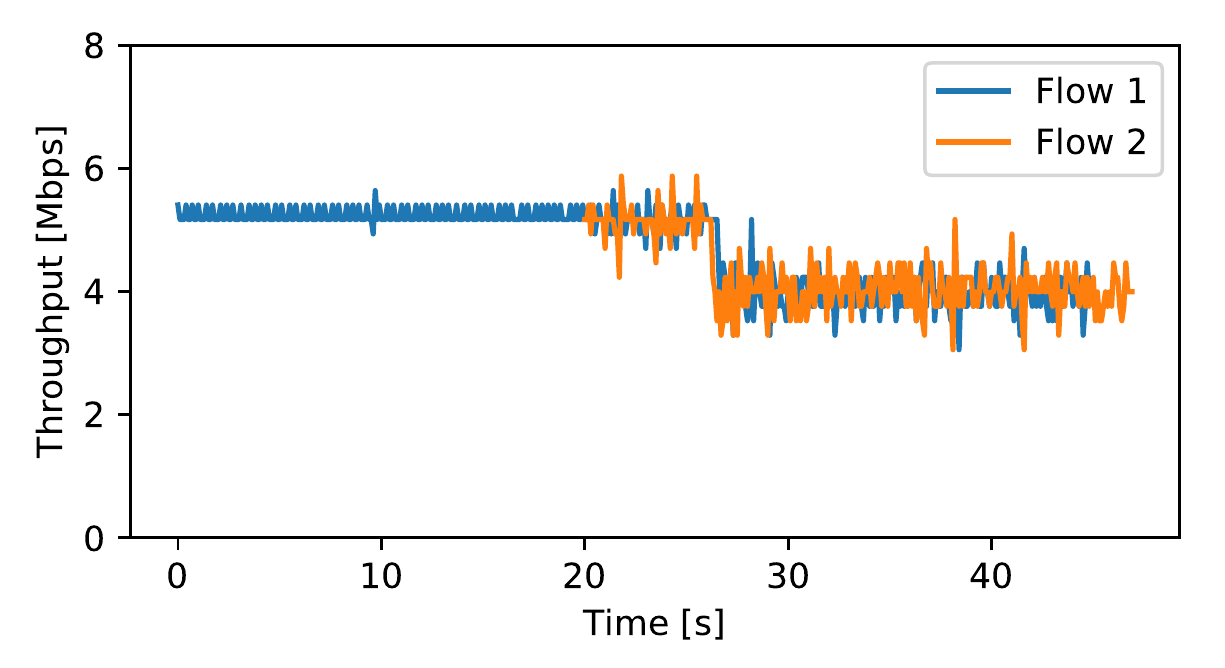}
	\caption{Distributed rate limiter with two flows at different edges of the network.}
	\label{fig:rl_p4}
\end{figure}

In Fig.~\ref{fig:rl_p4} we show an example of the distributed rate limiting  application in action. We create two flows: Flow 1 from AS 1 directed towards server cluster 1 and another flow from AS 3 directed towards server cluster 3. We consider shortest path routing and place state replicas in SW1 and SW3.
Flow 1 starts at time 0 with a rate of 5~Mbps while flow 2 starts with an offset of 20~s and with the same rate.
Although the flows do not cross each other at any point in the network, when the flow 2 starts both of them are rate limited to a predefined aggregate 8~Mbps threshold. Note that oscillations in throughput are due to the adopted probabilistic dropping scheme.

\subsubsection{Link-aware load balancing}

In~\cite{alizadeh2014conga} the authors propose a load balancing scheme for data center networks, based on the congestion level of individual links from the source leaf switch to the destination leaf switch. Source leaf switches keep track of local uplink congestion and of the downlink congestion from each spine switch to the destination leaf. When a new flow starts, the source leaf switch selects a path to the destination by considering the one that minimizes the maximum congestion on the whole path, i.e., local uplink congestion and the downlink congestion on the spine.

For the sake of simplicity, we present a reduced version of the application with some omitted details and by assuming that the application targets a single leaf switch with $P$ spine switches. The application can be easily extended to many leaf switches by simply instantiating multiple instances of the same application and  %Note that this case falls into the case of intra-application state sharing as 
the states related to downlink congestion must be shared across multiple leaf switches.

This network application can be mapped to LOADER as follows:

\begin{itemize}
	\item \textbf{States}: Given a leaf switch, we define state $s_i$ as the average load on the $P$ uplink ports, with $i=1,\ldots,P$. Additionally, we define state $s_j$ as the average downlink load on the port leading to the destination leaf switch of spine switch $j-P$,  with $j=P+1,\ldots,2P$.
	
	\item \textbf{Reduction function}: The reduction function is composed of two primitive actions, namely $r_1 = \max()$ and $r_2 = \arg\min()$. Consequently, the reduced version of the states is obtained as: $s^o=\arg\min(\max(s_1, s_{P+1}),\ldots,\max(s_P, s_{2P}))$

	\item \textbf{Trigger function}: Differently from previous use cases, the trigger function in this network application triggers the activity function each time a new $s^o$ is obtained and does not require any additional checks.

	\item \textbf{Activity function}: The activity function involves simple insertion of a new per-flow forwarding rule for each new flow based on the outcome of the reduction function.
\end{itemize}

% \subsubsection{Resource-aware load balancing}
% A resource-aware load balancing application has been introduced in Sec.~\ref{sec:application-definition}. The application performs load balancing of the user requests among the available servers based on the amount of available resources (e.g., average CPU utilization) at each server. 
% As we already defined the function mapping, in the following we present solely the code of the application. 
% For simplicity we do not define the states as they are not directly related to the network conditions, but instead to servers status. 

%% file: conclusion.tex
%!TEX root = main.tex
%!TEX spellcheck = en_US

\section{Conclusion}\label{sec:conc}
We propose a novel framework, namely LOADER, to address the limitation of stateful data planes in the  presence of non-local states at the switches in the definition of the network applications. LOADER enables stateful switches to take decisions based on information which is not locally available. This is achieved by introducing a state replication mechanism among the switches.
We discuss the main practical design challenges to support state replication, whose implementation is validated using both P4 and OPP stateful data planes.   

Furthermore, we provide a high-level programming abstraction for the development of distributed network applications based on replicated states.
Our programming model combines the expressiveness of a high-level programming model without ignoring the underlying hardware architecture of programmable switches. Thus, it is both of easy understanding for the programmer and can provide a comprehensible abstraction for the compilation and the embedding of network applications.

By combining the proposed abstraction model with the implementation of the replication mechanism, LOADER effectively permits to support distributed network-wide applications without involving any central entity.
As our results show, distributed network applications can be beneficial for the network performance and can be efficiently implemented in high-performance programmable stateful switches.

%% file: appendix_codes.tex
%!TEX root = main.tex
\appendix
\section{LOADER implementation of reference applications}\label{appendix:codes}
\subsection{DDoS detection with LOADER}
The DDoS detection application operates on a series of network states related to the transit SYN packets on the edge routers of the network (line 20). To filter the corresponding ports and the type of packets the \verb|scope| attribute is used during the definition of the state which is passed a helper function, namely \verb|extPortFilter|. The reduction function is simply defined as the sum of the states (line 26) through a predefined primitive. The trigger function (line 32) simply compares the outcome of the reduction function against a predefined threshold \verb|R| and invokes the activity function whenever the condition is satisfied. The inconsistency level is defined as time obsolescence with $\epsilon_t = 0.2$ms (line 24) forcing state replication to occur every 0.2ms.
The activity function targets all edge routers and acts by performing the notification of the controller about the presence of an attack. 

\begin{figure}[t]
    %\begin{center}
    %\begin{minipage}{0.9\columnwidth}\centering
\begin{lstlisting}[numbers=left, linewidth=0.95\columnwidth, language=Python2, caption=DDoS detection  with LOADER, label=lst:ddos]
from Controller import TopologyManager
from LOADER.PrimitiveActions import Drop, StateSum, Rate
from LOADER.Scope import Pkt

def extPortFilter(devices):
    extPorts = []
    for d in devices:
        extPorts += [p for p in d.getPorts() if p.Type==EXTERNAL]
    return (Pkt.ingressPort in extPorts) and (Pkt.TCP.Flag.SYN == 1)

R = 1000 # DDoS threshold in SYN pkts / s

# List of all edge routers
devices = TopologyManager.getEdgeRouters() 
applicationStates = []

# Iterate over all edge routers
for i in range(devices): 
    # Create a state for each edge router
    s = State(target=d, 
            scope=Rate(filter=Pkt(filter = extPortFilter([d]))))
    applicationStates.append(s)

# Define the reduction function as the sum of application states
r = ReductionFunction(states=applicationStates, 
        operation=StateSum)

# Define the activity function to drop all incoming packets
a = ActivityFunction(target=devices, scope=Pkt(filter=extPortFilter(devices)), action=Controller.Notify("DDoS detected"))

# Define trigger function to perform probabilistic dropping
tr = TriggerFunction(s0=r.Result(), 
            trigger=r.Result()>R,
            inconsistencyLevel=TimeObsolescence(0.2, "ms")
            activity = a)

\end{lstlisting}\squeezeup
\end{figure}

\subsection{Distributed rate limiting with LOADER}
The distributed rate limiting application is a variation of the DDoS application. Notably, while the reduction function remains invariant to the DDoS case, the states are defined as the total amount of traffic going through the edge routers (line 20). To perform rate limiting, the activity function (line 29) is defined to drop any incoming packet following the activation of the trigger function. The trigger function is randomly activated, with the probability of activating increasing whenever the total incoming traffic approaches the predefined target threshold \verb|R|.
As in the previous case the inconsistency level is defined as time obsolescence with $\epsilon_t = 0.2$ms.

\begin{figure}[h]
    %	\begin{center}
    %	\begin{minipage}{0.9\columnwidth}\centering
\begin{lstlisting}[numbers=left, linewidth=0.95\columnwidth, language=Python2, caption=Distributed rate limiting with LOADER, label=lst:drl]
from Controller import TopologyManager
from LOADER.PrimitiveActions import Drop, StateSum, Rate
from LOADER.Scope import Pkt

def extPortFilter(devices):
extPorts = []
for d in devices:
    extPorts += [p for p in d.getPorts() if p.Type==EXTERNAL]
return Pkt.ingressPort in extPorts

R = 100**6 # Desired rate in bps

# List of all edge routers
devices = TopologyManager.getEdgeRouters() 
applicationStates = []

# Iterate over all edge routers
for d in devices: 
# Create a state for each edge router
s = State(target=d, 
        scope=Rate(filter=Pkt(filter=extPortFilter([d]))))
applicationStates.append(s)

# Define the reduction function as the sum of application states
r = ReductionFunction(states=applicationStates, 
        operation=StateSum)

# Define the activity function to drop all incoming packets
a = ActivityFunction(target=devices,
        scope=Pkt(filter=extPortFilter(devices)), 
        action=Drop)

# Define trigger function to perform probabilistic dropping
tr = TriggerFunction(s0=r.Result(), 
        trigger=(rand()<(r.Result()-R)/r.Result()), 
        inconsistencyLevel=UpdateError(10),
        activity = a)
    
\end{lstlisting}\squeezeup
    %\end{minipage}
    %\end{center}
    \end{figure}

\subsection{Link-aware load balancing with LOADER}

In link-aware load balancing for data center networks, the objective is to find the least congested path from a given source server to a destination one. The states are defined as the congestion level on all uplink paths (line 25) and on the corresponding downlink paths (line 31). The uplink paths are considered by taking into account the set of leaf switches (line 18), while the downlink paths are taken over the spine switches (line 19). We assume that the topology manager exposes the appropriate methods to access the set of those switches. The reduction function performs a minmax operation over all the possible paths, thus leading to a path that minimizes the maximum congestion on the uplink-downlink segment (lines 14-16). This kind of application presents a trigger function which always returns true (lines 46-49). Consequently, the activity function is always triggered. Since the scope of the activity function targets all SYN packets (line 40), the activity is executed whenever a new flow arrives. Whenever this condition occurs, the activity function sets a new flow entry by assigning the least congested path for the new flow (lines 41-44). Notably, in the scenario of data center networks the load dynamics may change rapidly due to the presence of a big amount of flows. For this reason the inconsistency level is specified in the form of update error with $\epsilon_r=10$ writes (line 48), leading to a more updated information at the cost of potentially bigger synchronization traffic.

\begin{figure}[t]

\begin{lstlisting}[numbers=left, linewidth=0.9\columnwidth, language=Python2, caption=Link-aware load balancing with LOADER, label=lst:link-lb]
from Controller import TopologyManager
from LOADER.PrimitiveActions import SetEgress, Rate, min, max
from LOADER.Scope import Pkt

# Filter for downlink ports (i.e. from spine to leaf)
def dlPortFilter(device):
    return Pkt.getEgressPort() in [p for p in device.getPorts() if p.Type == DOWNLINK]

# Filter for uplink ports (i.e. from leaf to spine)
def ulPortFilter(device):
    return Pkt.getEgressPort() in [p for p in device.getPorts() if p.Type == UPLINK]

# Reduction function for minimum path congestion
def minMaxCong(ulCong, dlCong):
    dstLeaf = TopologyManager.getSpineID(Pkt.getDst())
    return argmin([max(ulCong[i], dlCong[i][dstLeaf]) for i in range(len(TopologyManager.getSpines()))])

l = TopologyManager.getLeafSwitches()[0]
spines = TopologyManager.getSpineSwitches()

dlCong = []
ulCong = []

for p in l.getPorts(filter = ulPortFilter):
    s = State(target=l, scope=Rate(filter = Port(p)))
    ulCong.append(s)

for sp in spines:
    spineLoad = []
    for p in sp.getPorts(filter = dlPortFilter):
        s = State(target=sp, scope=Rate(filter = p))	
        spineLoad.append(s)
    dlCong.append(spineLoad)

r = ReductionFunction(states=[ulCong, dlCong], 
            operation=minMaxCong)

a = ActivityFunction(
            target = l,
            scope = Pkt(filter = (Pkt.TCP.Flag.SYN == 1)), 
            action = insertRule(
            match = Pkt.getTuple(), 
            action = SetEgress,
            args = r.Result()))

tr = TriggerFunction(
            s0=r.Result(), 
            inconsistencyLevel=UpdateError(10),
            activity = a)
\end{lstlisting}\squeezeup

    %\end{minipage}
    %\end{center}
    \end{figure}